\documentclass[11pt]{article}
\renewcommand{\theequation}{\thesection.\arabic{equation}}
\newcounter{subequation}[equation]
\makeatletter

\expandafter\let\expandafter\reset@font\csname reset@font\endcsname

\def\subeqnarray{\arraycolsep1pt
    \def\@eqnnum\stepcounter##1{\stepcounter{subequation}%
	{\reset@font\rm(\theequation\alph{subequation})}}
\jot5mm     \eqnarray}

\makeatother

\def\tr{\mathop{\hbox{\rm tr}}\nolimits}
\def\str{\mathop{\hbox{\rm str}}\nolimits}

\def\u{\hbox{\rm u}}

\def\eref#1{(\ref{#1})}
\def\be{\begin{equation}}
\def\ee{\end{equation}}
\def\bea{\begin{eqnarray}}
\def\eea{\end{eqnarray}}

\def\half{\frac{1}{2}}

\def\one#1{#1^{\raise5pt\hbox{$\scriptstyle\!\!\!\!1$}}\,{}}
\def\two#1{#1^{\raise5pt\hbox{$\scriptstyle\!\!\!\!2$}}\,{}}

\makeatletter
\def\binrel@#1{\begingroup
  \setboxz@h{\thinmuskip0mu
    \medmuskip\m@ne mu\thickmuskip\@ne mu
    \setbox\tw@\hbox{$#1\m@th$}\kern-\wd\tw@
    ${}#1{}\m@th$}%
  \edef\@tempa{\endgroup\let\noexpand\binrel@@
    \ifdim\wdz@<\z@ \mathbin
    \else\ifdim\wdz@>\z@ \mathrel
    \else \relax\fi\fi}%
  \@tempa
}
\let\binrel@@\relax
\def\overset#1#2{\binrel@{#2}%
  \binrel@@{\mathop{\kern\z@#2}\limits^{#1}}}
\def\underset#1#2{\binrel@{#2}%
  \binrel@@{\mathop{\kern\z@#2}\limits_{#1}}}
\makeatother
\newfont{\bbd}{msbm10 scaled\magstep1}

\def\R{\hbox{\bbd R}}
\def\T{\hbox{\bbd T}}

\def\P{\hbox{\bbd P}}

\def\E{{\mathcal E}}
\def\H{{\mathcal H}}

\def\Z{{\mathcal Z}}

\def\TT{{\mathcal T}}

\begin{document}
\hfill NTZ 28/1999

{\begin{center}
{\LARGE {Heisenberg spin chains based on 
$s\ell(2|1)$ symmetry.} } \\ [8mm]
{\large  S.Derkachov$^{a,c}$\footnote{e-mail: 
Sergey.Derkachov@itp.uni-leipzig.de}, 
D. Karakhanyan$^b$\footnote{e-mail: karakhan@lx2.yerphi.am} \& 
R. Kirschner$^c$\footnote{e-mail:Roland.Kirschner@itp.uni-leipzig.de} \\ [3mm] } 
\end{center} 
} 

\begin{itemize}
\item[$^a$] 
Department of Mathematics, St Petersburg Technology Institute, \\
Sankt Petersburg, Russia   
\item[$^b$] 
Yerevan Physics Institute , \\
Br.Alikhanian st.2, 375036, Yerevan , Armenia. 
\item[$^c$] Naturwissenschaftlich-Theoretisches Zentrum und 
Institut f\"{u}r Theoretische Physik, Universit\"{a}t Leipzig, \\
Augustusplatz 10, D-04109 Leipzig, Germany
\end{itemize}

\vspace{3cm} 
\noindent 
{\bf Abstract.}   
We find solutions of the Yang-Baxter 
equation acting on tensor product of 
arbitrary representations of the superalgebra
$s\ell(2|1)$. 
Based on these solutions we construct the local Hamiltonians 
for integrable homogeneous periodic chains and open chains.

\newpage


{\small \tableofcontents}
\renewcommand{\refname}{References.}
\renewcommand{\thefootnote}{\arabic{footnote}}
\setcounter{footnote}{0} 
\setcounter{equation}{0}

\renewcommand{\theequation}{\thesubsection.\arabic{equation}}
\setcounter{equation}{0}

\section{Introduction}
\setcounter{equation}{0}

The superalgebra~$s\ell(2|1)$~\cite{SNR},~\cite{JG} appears 
in various quantum systems as underlying their symmetry and dynamics.
Finite-dimensional representations describe spin states. 
For example, a lattice site which is allowed to 
be empty or occupied by an electron with the two spin 
states $\pm\half$ , but not by two electrons, 
corresponds to the three-dimensional fundamental representation.
Chains consisting of sites carrying the fundamental 
representation with integrable short-range 
interaction have been constructed.  
The Hamiltonian of $t-J$ model is obtained 
from the transfer matrix of the integrable model 
based on the three dimensional fundamental 
representation of~$s\ell(2|1)$~\cite{EK},~\cite{FK}.  
The superalgebra~$s\ell(2|1)$ has a series of four-dimensional 
representations parametrized by a parameter $b\ne \pm\half$.
Integrable models built from R-matrices defined on 
tensor products of two different four-dimensional 
representations have been 
considered in~\cite{Maas},~\cite{RM},~\cite{PF}.\\   
The simplest integrable chain structure is the 
homogeneous periodic spin chain; 
most applications make use of this case.
There exist some important modifications.\\
The construction of open spin chains is well known~\cite{S}.
The treatment of integrable inhomogeneous 
spin chains is more involved.
An integrable model has been constructed~\cite{AJ}
in view of the relevance for systems with impurities 
and in particular for the Kondo effect.
The representations of~$s\ell(2|1)$ accommodate~$s\ell(2)$ 
representations of spin $s$ and $s\pm\half$ and 
allow in this way to construct chains with 
mobile impurities~\cite{FPT}.\\  
The integrable chains turn out to describe approximately 
the effective interaction in four-dimensional gauge theories 
in the Regge limit~\cite{L},~\cite{FK} 
and in the Bjorken limit~\cite{BDM}.
Unlike the above examples here the sites carry 
infinite-dimensional representations of~$s\ell(2)$ accommodating 
all the momentum states of reggeons and partons.
Besides of the case of homogeneous periodic chains 
also the case of open chains is encountered 
both in the Bjorken limit~\cite{DKM,B} 
and in the Regge limit~\cite{KK}.  
A particular~$s\ell(2|1)$ representation of interest are the 
infinitesimal conformal transformations in one 
dimension together with their twofold supersymmetric extensions.
This symmetry applies to the Bjorken limit of 
four-dimensional supersymmetric Yang-Mills 
theory at least up to one loop~\cite{BFLK},~\cite{BM}.
This means, the one-loop renormalization 
of quasipartonic composite operators can 
be obtained by~$s\ell(2|1)$ symmetric pairwise 
interactions of partons, the 
states~(light-cone momenta, helicity,fermion number) of which 
form an infinite-dimensional lowest weight module of this algebra.\\
In the present paper we consider the algebra~$s\ell(2|1)$, 
its lowest weight modules and construct on this basis the 
solutions of the Yang-Baxter equation, 
i.e. the R-matrix acting on the tensor product of two those modules.
R-matrices acting on tensor products of two 
fundamental~$s\ell(2|1)$-representations and of 
two four-dimensional representations have been 
constructed in the above mentioned 
papers~\cite{EK},~\cite{Maas},~\cite{RM},~\cite{PF}. 
General integrable models based on R-matrices 
acting on tensor product of two arbitrary 
finite-dimensional atypical(chiral) representations of~$s\ell(2|1)$ 
have been constructed quite recently~\cite{F} by 
generalizing the known approach~\cite{KRS} 
from~$s\ell(2)$ to the case of~$s\ell(2|1)$.\\  
We propose the alternative approach and generalize these results.
Motivated by possible applications to the Bjorken limit in QCD, 
we represent the lowest weight modules by polynomials in one even~($z$)
and two odd~($\theta,\bar\theta$) variables.
The general R-matrices are in fact operators acting 
on two-point functions, i.e. (polynomial) functions of 
two sets~($z_1,\theta_1,\bar\theta_1$) and~($z_2,\theta_2,\bar\theta_2$) 
representing the tensor product. We construct the R-operators acting 
on the tensor product of two arbitrary~(finite or infinite-dimensional) 
~$s\ell(2|1)$-modules.
This is done by calculating the two-point eigenfunctions of the 
lowest weight and the eigenvalues of the R-operator.\\
From the particular result for arbitrary but isomorphic 
representations we derive the integrable nearest neighbour 
interaction Hamiltonian for homogeneous periodic chains 
with sites carrying arbitrary isomorphic representations.\\
In the case of~$s\ell(2)$ there exist integrable nearest neighbour    
interactions in open chains homogeneous inside but with arbitrary 
different representations corresponding to the end points~\cite{DKM}.
We extend this result to the case of~$s\ell(2|1)$ and 
construct the corresponding Hamiltonians.\\    
The presentation is organized as follows.
In Section 2 we introduce definitions and 
summarize the standard facts 
about the superalgebra~$s\ell(2|1)$ and its 
representations.
We represent the lowest weight modules by polynomials in one 
even~($z$) and two odd variables~($\theta,\bar\theta$) and 
the $s\ell(2|1)$-generators as first order 
differential operators.\\
In Section 3 we derive the defining relation for 
the general R-matrix, i.e. the solution of the Yang-Baxter equation 
acting on tensor products of two arbitrary representations, 
the elements of which are polynomial functions 
of~($z_1,\theta_1,\bar\theta_1$) and~($z_2,\theta_2,\bar\theta_2$). 
We solve this defining relation in the space of lowest weights.\\
In Section 4 we construct the local integrable Hamiltonians 
in the simplest case of homogeneous periodic chain 
carrying arbitrary isomorphic representations on the sites.\\
In Section 5 we construct the local integrable Hamiltonians 
for the open chain with arbitrary isomorphic representations inside 
and other arbitrary representations at the end points.

Finally, in Section 6 we summarize.
Appendix A contains some technical details of calculations.
In Appendix B we give the expression for 
the R-matrix acting in the tensor product of chiral modules.
In Appendix C we discuss shortly the 
case of finite-dimensional representations and show that 
obtained general R-matrix reduces 
to the known ones~\cite{PF} for 
the tensor product of modules with minimal dimensions.

\section{Algebra $s\ell(2|1)$}
\setcounter{equation}{0}
\label{sec:sl21}

\subsection{Commutators and Casimir Operators}
\setcounter{equation}{0}

The superalgebra $s\ell(2|1)$ has eight generators:
four odd $V^{\pm},W^{\pm}$
and four even ones $S,S^{\pm}$ and $B$.
The commutation relations have the following form~\cite{SNR}:
anticommutators
$$
\{V^{\pm},V^{\pm}\} =0\ ;\  \{V^{\pm},V^{\mp}\} =0 
\ ;\ \{W^{\pm},W^{\pm}\} =0\ ;\  \{W^{\pm},W^{\mp}\} =0 
$$
$$
\{V^{\pm},W^{\pm}\} =\pm S^{\pm}
\ ;\  \{V^{\pm},W^{\mp}\} = - S \pm B,
$$
commutators
$$
[S,S^{\pm}] =\pm S^{\pm}\ ;\  [S^{+},S^{-}] = 2 S 
$$
$$
[B,S^{\pm}] = 0\ ;\  [B,S] = 0
\ ;\  [S^{\pm},V^{\pm}] = 0\ ;\  [S^{\pm},W^{\pm}] = 0
$$
$$
[B,V^{\pm}] = \half V^{\pm}\ ;\  [B,W^{\pm}] = -\half W^{\pm}
\ ;\ [S,V^{\pm}] = \pm \half V^{\pm}
\ ;\  [S,W^{\pm}] = \pm \half W^{\pm}
$$
$$
[S^{\pm},V^{\mp}] = V^{\pm}\ ;\  [S^{\pm},W^{\mp}] = W^{\pm}.
$$
These generators are linear combinations of 
the generators $E_{AB}$ of 
the superalgebra $g\ell(2|1)$~\cite{JG}. 
The commutation relations for the nine 
generators of $g\ell(2|1)$ 
can be written compactly in the form:
$$
[E_{AB},E_{CD}] = 
\delta_{CB} E_{AD} - 
(-)^{(\bar{A}+\bar{B})(\bar{C}+\bar{D})}
\delta_{AD} E_{CB}
$$
where the graded commutator is defined as:
$$
[E_{AB},E_{CD}]\equiv 
E_{AB}\cdot E_{CD} - 
(-)^{(\bar{A}+\bar{B})(\bar{C}+\bar{D})}
E_{CD}\cdot E_{AB}
$$
The indices $A,B,C,D = 1,2,3$ and we choose 
the grading:$\bar 1 =\bar 3 =0\ ;\ \bar 2 = 1$.
The connection between both sets of generators is the following:
$$
S^{-} = E_{31}\ ;\ W^{-} = -E_{21}\ ;\ V^{-} = E_{32} 
$$
$$
S^{+} = E_{13}\ ;\ W^{+} = E_{23}\ ;\ V^{+} = E_{12} 
$$
\be
S = \half E_{11}- \half E_{33}
\ ;\  B = -\half E_{11}-E_{22}-\half E_{33} 
\label{sl2}
\ee
In the fundamental representation all generators $e_{AB}$ of $g\ell(2|1)$ 
are $3\times 3$-matrices and the basis in the space of these matrices 
can be chosen in the standard way:
\be
(e_{AB})_{CD} = \delta_{AC}\delta_{BD}
\ ;\  e_{AB}e_{CD} = \delta_{CB}e_{AD}
\label{ee}
\ee
In the fundamental representation 
the generators $W^{\pm},V^{\pm},S^{\pm},S,B$ 
have the form:  
$$
S^{-} = 
\left (\begin{array}{ccc}
0 & 0 & 0  \\
0 & 0 & 0  \\ 
1 & 0 & 0    
\end{array} \right )
\ ;\  W^{-} = 
\left (\begin{array}{ccc}
0 & 0 & 0  \\
-1 & 0 & 0  \\ 
0 & 0 & 0    
\end{array} \right )
\ ;\  V^{-} = 
\left (\begin{array}{ccc}
0 & 0 & 0  \\
0 & 0 & 0  \\ 
0 & 1 & 0    
\end{array} \right )
$$
$$
S^{+} = 
\left (\begin{array}{ccc}
0 & 0 & 1  \\
0 & 0 & 0  \\ 
0 & 0 & 0    
\end{array} \right )
\ ;\  W^{+} = 
\left (\begin{array}{ccc}
0 & 0 & 0  \\
0 & 0 & 1  \\ 
0 & 0 & 0    
\end{array} \right )
\ ;\  V^{+} = 
\left (\begin{array}{ccc}
0 & 1 & 0  \\
0 & 0 & 0  \\ 
0 & 0 & 0    
\end{array} \right )
$$
\be
S = 
\left (\begin{array}{ccc}
\half & 0 & 0  \\
0 & 0 & 0  \\ 
0 & 0 & -\half    
\end{array} \right )
\ ;\  B = 
\left (\begin{array}{ccc}
-\half & 0 & 0  \\
0 & -1 & 0  \\ 
0 & 0 & -\half    
\end{array} \right )
\label{fun}
\ee
There exists a simple construction for the 
central elements of the enveloping algebra of $g\ell(2|1)$~\cite{JG}.
The first step is the construction of covariant operators:
suppose we have two covariant operators $V^{(i)}_{CD}$, 
i.e. operators which have the following 
commutation relations with generators
$$
[E_{AB},V^{(i)}_{CD}] = 
\delta_{CB} V^{(i)}_{AD} - 
(-)^{(\bar{A}+\bar{B})(\bar{C}+\bar{D})}
\delta_{AD} V^{(i)}_{CB}\ ;\ i=1,2.
$$
It is easy to check that operator 
$V_{AB} = \sum_C (-)^{\bar C} V^{(1)}_{AC}V^{(2)}_{CB}$
is also covariant.
This simple observation allows to construct 
covariant operators using generators $E_{CD}$ 
as simplest building blocks.
The second step is the construction of a central 
element from the covariant operator:
for any covariant operator $V_{CD}$ 
the operator $V = \sum_{C}V_{CC}$ belongs 
to the center of the algebra
$$
[E_{AB}, V] = 0.
$$
Repeating this construction we obtain central 
elements $K_n,~n=1,2,3...$~\cite{DIC} for the enveloping 
algebra of $g\ell(2|1)$: 
\be
K_1 = \sum_{A}E_{AA}
\ ;\ K_2 = \sum_{AB} (-)^{\bar B} E_{AB}E_{BA}
\ ;\ K_3 = \sum_{ABC} (-)^{\bar B +\bar C} E_{AB}E_{BC}E_{CA}  
\ ;\ ...
\label{K}
\ee 
The eight generators of algebra $s\ell(2|1)$ may be introduced by defining:
\be
\E_{AB}\equiv E_{AB} - \delta_{AB}(-)^{\bar B}\sum_{A}E_{AA}
\label{sl}
\ee
$$
\E_{31} = S^{-}\ ;\ \E_{21} = - W^{-}\ ;\ \E_{32} = V^{-} 
$$
$$
\E_{13} = S^{+}\ ;\ \E_{23} = W^{+}\ ;\ \E_{12} = V^{+}  
$$
$$
\E_{11} = B - S\ ;\ \E_{22} = -2 B\ ;\  \E_{33} = B + S.
$$
It can be verified that generators $\E_{AB}$ satisfy the same 
commutation relations as $E_{AB}$:
$$
[\E_{AB},\E_{CD}] = 
\delta_{CB} \E_{AD} - 
(-)^{(\bar{A}+\bar{B})(\bar{C}+\bar{D})}
\delta_{AD} \E_{CB}
$$
There exists only one restriction for the Cartan generators:
$\E_{11}+\E_{22}+\E_{33} = 0$ and the independent 
eight generators can be chosen in the form~(\ref{sl2}).\\ 
The center of algebra $s\ell(2|1)$ is generated  
by Casimir operators~$C_n,~n=2,3...$~\cite{DIC}.
We shall use only two of them:
\be
C_2 = \frac{1}{2!}\sum_{AB} (-)^{\bar B} \E_{AB}\E_{BA} = 
S^2-B^2+S^{+}S^{-}+V^{+}W^{-}+W^{+}V^{-}  
\label{C_2}
\ee  
\be
C_3 =\frac{1}{3!} \sum_{ABC} (-)^{\bar B +\bar C} \E_{AB}\E_{BC}\E_{CA}= 
B(S^2-B^2) + B S^{+}S^{-}+ 
\frac{3}{2}B (V^{+}W^{-}+W^{+}V^{-})+
\label{C_3}
\ee
$$
+\frac{1}{4}(W^{+}V^{+}-V^{+}W^{+})S^{-}+
\frac{1}{4}S^{+}(V^{-}W^{-}-W^{-}V^{-})+
\frac{1}{2}(S-1)(V^{+}W^{-}-W^{+}V^{-}).
$$

\subsection{Global form of superconformal transformations}
\setcounter{equation}{0}

We represent the generators as first order differential operators,
acting on the space of polynomials $\Phi(z,\theta,\bar\theta)$. 
Lowering(decreasing the polynomial degree) operators have the form   
\be
S^{-} = -\partial
\ ;\ V^{-} = \partial_{\theta}+\frac{1}{2}\bar\theta\partial
\ ;\ W^{-} = \partial_{\bar\theta}+\frac{1}{2}\theta\partial
\label{gen-}
\ee
and generate the following global transformations
\be
e^{\lambda S^{-}}\Phi(z;\theta,\bar\theta)=
\Phi(z-\lambda;\theta,\bar\theta),
\label{glob-}
\ee
$$
e^{\epsilon V^{-}}\Phi(z;\theta,\bar\theta)=
\Phi\left(z+\frac{\epsilon\bar\theta}{2};\theta+\epsilon,\bar\theta\right)
\ ;\ e^{\epsilon W^{-}}\Phi(z;\theta,\bar\theta)=
\Phi\left(z+\frac{\epsilon\theta}{2};\theta,\bar\theta+\epsilon\right).
$$
Rising(increasing the polynomial degree) operators
$$
V^{+}= -\left[z\partial_{\theta} +\frac{1}{2}\bar\theta z\partial+
\frac{1}{2}\bar\theta\theta\partial_{\theta}\right]-(\ell-b)\bar\theta
\ ;\ W^{+}= -\left[z\partial_{\bar\theta} +\frac{1}{2}\theta z\partial+
\frac{1}{2}\theta\bar\theta\partial_{\bar\theta}\right]-(\ell+b)\theta ,
$$
\be
S^{+}= z^2\partial + z\theta\partial_{\theta}+ 
z\bar\theta\partial_{\bar\theta} + 2\ell z 
-b\theta\bar\theta , 
\label{gen+}
\ee 
generate the global transformations
$$
e^{\lambda S^{+}}\Phi(z;\theta,\bar\theta)=
\left[1+\frac{\theta\bar\theta\lambda}
{(1-\lambda z)}\right]^{-b}
\frac{1}{(1-\lambda z)^{2\ell}}
\Phi\left(\frac{z}{1-\lambda z};\frac{\theta}{1-\lambda z},
\frac{\bar\theta}{1-\lambda z}\right) ,
$$
$$
e^{\epsilon V^{+}}\Phi(z;\theta,\bar\theta)=
\frac{1}{(1+\epsilon\bar\theta)^{\ell-b}}
\Phi\left(\frac{z}{1+\frac{\epsilon \bar\theta}{2}};
\frac{\theta -\epsilon z}
{1+\frac{\epsilon \bar\theta}{2}} ,\bar\theta\right) ,
$$
$$
e^{\epsilon W^{+}}\Phi(z;\theta,\bar\theta)=
\frac{1}{(1+\epsilon \theta)^{\ell+b}}
\Phi\left(\frac{z}{1+\frac{\epsilon \theta}{2}};\theta,
\frac{\bar\theta -\epsilon z}{1+\frac{\epsilon \theta}{2}}\right) .
$$
Two remaining elements of the Cartan subalgebra:
$$
S= z\partial +\frac{1}{2}\theta\partial_{\theta}+
\frac{1}{2}\bar\theta\partial_{\bar\theta} +\ell
\ ;\  B= \frac{1}{2}\bar\theta\partial_{\bar\theta} -
\frac{1}{2}\theta\partial_{\theta}+b
$$
generate the transformations:
$$
e^{\lambda S}\Phi(z;\theta,\bar\theta)=
e^{\ell\lambda}
\Phi\left(e^{\lambda}z;e^{\frac{\lambda}{2}}\theta,
e^{\frac{\lambda}{2}}\bar\theta\right)
$$
$$
e^{\lambda B}\Phi(z;\theta,\bar\theta)=
e^{b\lambda}
\Phi\left(z;e^{-\frac{\lambda}{2}}\theta,
e^{\frac{\lambda}{2}}\bar\theta\right)
$$
We use a natural convention here and assign 
scaling dimension one and $U(1)$-charge zero 
to the even variable $z$ and the scaling 
dimension $\half$ and $U(1)$-charge $\mp\half$ to the 
odd variables $\theta$ and $\bar\theta$.

\subsection{$s\ell(2|1)$-lowest weight modules}
\setcounter{equation}{0}

The lowest weight $s\ell(2|1)$-module 
$V_{\ell,b} = V_{\vec\ell}\ ,\ \vec\ell = (\ell,b)$ 
is built on the lowest weight vector $\psi$ obeying:
$$
V_{-} \psi = 0\ ;\  W^{-}\psi = 0
\ ;\  S^{-}\psi = 0\ ;\ S \psi = \ell\psi
\ ;\  B\psi = b \psi
$$
In generic situation $\ell\ne\pm b$ the module 
is characterised uniquely by the action of 
the Casimir operators on its elements:
$$
C_2 v = (\ell^2-b^2) v
\ ;\  C_3 v = b(\ell^2-b^2) v 
\ ;\ v \in V_{\ell,b} 
$$
It is a vector space spanned by the following 
basis~\cite{SNR},~\cite{Mar} with the even vectors
$$
A_k = (S^{+})^k \psi \ ;\  B_k = (S^{+})^{k-1}W^{+}V^{+}\psi
\ ,\  k\in \Z_{+}
$$
and the odd vectors
$$
V_k = (S^{+})^k V^{+}\psi \ ;\  W_k = (S^{+})^{k}W^{+}\psi
\ ,\  k\in \Z_{+}
$$
We shall use the above realization of the 
$s\ell(2|1)$-generators as the differential 
operators of first order acting on the
infinite-dimensional(for generic $\ell$) space 
$V_{\ell,b}$ of polynomials~$\Phi(z,\theta,\bar\theta)$ 
of variables $z,\theta,\bar\theta$.\\
It is easy to obtain the expression for 
the coherent states 
$e^{\lambda S^{+}}\psi\ , \ e^{\lambda S^{+}}W^{+}V^{+}\psi$ and
$e^{\lambda S^{+}}V^{+}\psi\ , \ e^{\lambda S^{+}}W^{+}\psi$ 
for the lowest weight $\psi = 1$ using 
the formulae of the previous section.
They are the generating functions , the power expansion in 
$\lambda$ of which produces the basis: 
\be
A_{k}=(2\ell)_k \left[z^k-
\frac{k b}{2\ell}z^{k-1}\theta\bar\theta\right] 
\ ;\  B_{k}= \frac{\ell-b}{2\ell}(2\ell)_k\left[z^k+
\left(b+\ell+\frac{k}{2}\right)
z^{k-1}\theta\bar\theta\right] 
\label{AB}
\ee
$$
V_{k}=-(\ell-b)(2\ell+1)_k z^k\bar\theta 
\ ;\ W_{k}=-(\ell+b)(2\ell+1)_k z^k\theta 
\ ;\ (2\ell)_k \equiv \frac{\Gamma(2\ell+k)}{\Gamma(2\ell)}
$$ 
It is useful to introduce the subspaces of 
functions with definite chirality.
Let us define for this purpose two operators 
called supercovariant derivatives:
\be
D^{+} = -\partial_{\theta}+\frac{1}{2}\bar\theta\partial
\ ;\ D^{-} = -\partial_{\bar\theta}+\frac{1}{2}\theta\partial
\label{cir}
\ee
and two subspaces $V^{\pm}_{\ell,b}\equiv \ker D^{\pm}\cap V_{\ell,b}$:
$$
\Phi(z,\theta,\bar\theta)\in V^{+}_{\ell,b} \Rightarrow 
\Phi(z,\theta,\bar\theta)= \Phi(z_{+},\theta_{+})
\ ;\ z_{+}\equiv z+\half \theta\bar\theta
\ ,\ \theta_{+}\equiv\bar\theta
$$
$$
\Phi(z,\theta,\bar\theta)\in V^{-}_{\ell,b} \Rightarrow 
\Phi(z,\theta,\bar\theta)= \Phi(z_{-},\theta_{-})
\ ;\ z_{-}\equiv z-\half \theta\bar\theta
\ ,\ \theta_{-}\equiv \theta
$$
In the generic case the chiral subspaces $V^{\pm}_{\ell,b}$
are not $s\ell(2|1)$-invariant ones.
Indeed, the operators $D^{\pm}$ have the 
following commutation relations with $s\ell(2|1)$-generators:
$$
\{D^{\pm},V^{-}\} = 0\ ;\  \{D^{\pm},W^{-}\} =0
\ ;\  [D^{\pm},S^{-}] = 0 
$$
\be
[D^{\pm},S] = \half D^{\pm}
\ ;\  [D^{\pm},B] = \mp \half D^{\pm}
\label{crD}
\ee
$$
\{D^{+},V^{+}\} =\bar\theta D^{+}
\ ;\  \{D^{+},W^{+}\} = \ell+b
\ ;\  [D^{+},S^{+}] = (\ell+b)\bar\theta + z_{+}D^{+}  
$$
$$
\{D^{-},V^{+}\} = \ell-b 
\ ;\  \{D^{-},W^{+}\} = \theta D^{-}
\ ;\  [D^{-},S^{+}] = (\ell-b)\theta + z_{-}D^{-}, 
$$
and it is easy to see that chiral subspaces 
$V^{\pm}_{\ell,b}$ are $s\ell(2|1)$-invariant 
only under the condition $\ell = \mp b$.
In this case the whole module 
$V_{\ell,\pm\ell}$ has definite chirality  
$V_{\ell,\pm\ell} = V^{\mp}_{\ell,\pm\ell}$:
$$
D^{-}v = 0\ ,\ v\in V_{\ell,\ell}
\ ;\ D^{+}v = 0\ ,\ v\in V_{\ell,-\ell}. 
$$
The notions of chirality and chiral 
representations are used here as they
are common in supersymmetric field theory.
In the mathematical literature about 
superalgebra representations 
generic representations are called typical and chiral 
representations are called atypical~\cite{SNR},~\cite{Mar}.
There exist some special values of $\ell$:
$\ell = - n\ ;\ n \in \half \Z_+$ , for which 
the module $V_{\ell,b}$ becomes a 
finite-dimensional vector space~\cite{SNR},~\cite{Mar}.
Indeed it is evident from~(\ref{AB}) that all basis vectors 
are equal zero for $k\geq n+1$.
There are three cases depending on the relation between $b$ and $n$.
The first case is for generic $b$(typical representations): 
$b \ne \pm n\ ;\ \dim V_{-n,b} = 8 n $
$$
\Phi^{\pm}_k = z_{\pm}^k\ ,\ k= 1...2 n -1
\ ,\ \Phi_0 = 1\ ,\ \Phi_{2 n} = 
\left(z+\frac{b}{2n}\theta\bar\theta\right)^{2 n}
\ ;\ \Psi^{\pm}_k = \theta_{\pm} z_{\pm}^{k}
\ ,\ k= 0...2 n-1 
$$
The second and third case appears for~$b = \pm n$ and 
here the representation spaces have definite chirality
(atypical representations).\\
$b = \pm n\ ;\ V_{-n,\pm n} = V^{\pm}_{-n,-n}
\ ;\ \dim V_{-n,\pm n} = 4 n+1$ 
$$
\Phi^{\pm}_k = z_{\pm}^k
\ ,\ k= 0...2 n
\ ;\ \Psi^{\pm}_k = \theta_{\pm}z_{\pm}^{k}\ ,\ k= 0...2 n - 1
$$
Let us introduce the special notation 
for the fundamental $s\ell(2|1)$-module:\\ 
$V_{-\half,-\half}~\equiv~V_{\vec f}\ ,\ \vec f =(-\half,-\half)$.
In the basis 
$$
e_1 = -z_{-} \leftrightarrow 
\left (\begin{array}{c}
1 \\ 0 \\0 \\   
\end{array} \right )
\ ;\ e_2 = \theta_{-} 
\leftrightarrow 
\left (\begin{array}{ccc}
0 \\1  \\ 0    
\end{array} \right )
\ ;\ e_3 = -1 \leftrightarrow 
\left (\begin{array}{ccc}
0 \\0  \\ 1    
\end{array} \right )
$$
the $s\ell(2|1)$-generators take their fundamental form~(\ref{fun}).

\subsection{Tensor products of two $s\ell(2|1)$-modules }
\setcounter{equation}{0}

The tensor product of two $s\ell(2|1)$-modules has 
the following direct sum decomposition~\cite{Mar}:
\be
V_{\ell_1,b_1}\otimes V_{\ell_2,b_2} = V_{\ell,b}+
2\sum_{n=1}^{\infty} V_{\ell+n,b}+
\sum_{n=0}^{\infty} V_{\ell+n+\half,b-\half}+
\sum_{n=0}^{\infty} V_{\ell+n+\half,b+\half}  
\ ;\ \ell_i\ne\pm b_i 
\label{sum}
\ee
$$
\ell = \ell_1+\ell_2\ ;\ b = b_1+b_2
$$
Note that this formula is applicable in the 
generic situation $\ell_i\ne\pm b_i$.
The direct sum decomposition 
reduces for the tensor product involving chiral modules:
$$
V_{\ell_1,\mp\ell_1}\otimes V_{\ell_2,b_2} = 
\sum_{n=0}^{\infty} V_{\ell+n,b}+
\sum_{n=0}^{\infty} V_{\ell+n+\half,b\mp\half}  
\ ;\ \ell_2\ne\pm b_2  
$$
$$
V_{\ell_1,\mp\ell_1}\otimes V_{\ell_2,\pm\ell_2} = 
\sum_{n=0}^{\infty} V_{\ell+n,b}
\ ;\  V_{\ell_1,\mp\ell_1}\otimes V_{\ell_2,\mp\ell_2} = 
\sum_{n=0}^{\infty} V_{\ell+n+\half,b\mp\half}  
$$
In Appendix C we discuss the modifications of~(\ref{sum}) 
arising for finite-dimensional representations~\cite{Mar}.\\
For the proof of~(\ref{sum}) one has to 
determine all possible lowest weight vectors 
appearing in the tensor 
product~$V_{\ell_1,b_1}\otimes V_{\ell_2,b_2}$.
In the realization on functions of $z,\theta,\bar\theta$ the space 
$V_{\ell_1,b_1}\otimes V_{\ell_2,b_2}$ is isomorphic 
to the space of polynomials on two even 
variables $z_1,z_2$ and four odd 
variables $\theta_1,\bar\theta_1,\theta_2,\bar\theta_2$ 
called for the sake of brevity two-point functions. 
The $s\ell(2|1)$-generators acting on the
$V_{\ell_1,b_1}\otimes V_{\ell_2,b_2}$ are just the sums 
of corresponding generators acting in $V_{\ell_i,b_i}$.
The lowest weight vectors of the irreducible 
representations in the decomposition of 
$V_{\ell_1,b_1}\otimes V_{\ell_2,b_2}$ are defined 
as the common solutions of the equations:
\be
S^{-}\Phi = V^{-}\Phi = W^{-}\Phi= 0
\label{lw}
\ee
which have the form
\be
\Phi(z_1,z_2;\theta_1,\theta_2;\bar\theta_1,\bar\theta_2) = 
\Phi\left(Z_{12}, \theta_{12}, \bar\theta_{12}\right)
\label{lwf}
\ee
where 
$$
Z_{12}\equiv 
z_1-z_2+\frac{1}{2}(\bar\theta_1\theta_2+\theta_1\bar\theta_2)
\ ;\ \theta_{12}\equiv \theta_1 - \theta_2
\ ;\ \bar\theta_{12}\equiv \bar\theta_1 - \bar\theta_2.
$$
Indeed, from~(\ref{lw}) follows immediately that the function 
$\Phi$ has to be invariant with respect 
to global transformations~(\ref{glob-}).
This invariance predicts the general form of~$\Phi$:
$$
e^{\beta W^{-}}e^{\alpha V^{-}}e^{a S^{-}}
\Phi(z_1,z_2;\theta_1,\theta_2;\bar\theta_1,\bar\theta_2)=
$$
$$
=\Phi\left(z_1-a+\frac{\alpha(\bar\theta_1+\beta)}{2}+\frac{\beta\theta_1}{2},
z_2-a+\frac{\alpha(\bar\theta_2+\beta)}{2}+\frac{\beta\theta_2}{2};
\theta_1+\alpha,\theta_2+\alpha;\bar\theta_1+\beta,\bar\theta_2+\beta\right)
$$
and choosing
$
a = z_2+\frac{1}{2}\theta_2\bar\theta_2 
\ ,\ \alpha = -\theta_2 \ ,\ \beta = -\bar\theta_2
$
we obtain~(\ref{lwf}).\\ 
There are additional restrictions of definite chirality 
for the lowest weights in the tensor product of the chiral modules:
$$
D^{\pm}_1\Phi = 0 \Rightarrow 
\Phi = \Phi\left(Z_{12}\pm\half \theta_{12}\bar\theta_{12},
\theta^{\pm}_{12}\right)
\ ;\  \theta^{\pm}_{12}\equiv\theta^{\pm}_{1}-\theta^{\pm}_{2} 
$$
$$
D^{\pm}_2\Phi = 0 \Rightarrow 
\Phi = \Phi\left(Z_{12}\mp\half \theta_{12}\bar\theta_{12},
\theta^{\pm}_{12}\right). 
$$
Now, the lowest weight vectors in the decomposition of 
the tensor product are constructed from functions~(\ref{lwf}) 
being eigenfunctions of generators $S$ and $B$.
The eigenfunctions of the operator 
$S$ are the polynomials with scaling dimension $n$ and 
the eigenfunctions of the operator 
$B$ are the polynomials with one of the possible 
$U(1)$-charges:~$ 0 , \pm\half$.\\
Finally we obtain that all lowest weights in the space 
$V_{\ell_1,b_1}\otimes V_{\ell_2,b_2}$ are divided on two sets,
the even lowest weights:
\be
\Phi^{\pm}_n\equiv 
\left(Z_{12}\pm\half\theta_{12}\bar\theta_{12}\right)^{n}
\ ;\ D_1^{\pm}\Phi^{\pm} = 0
\ ,\ S \Phi^{\pm}_n = (n +\ell)\Phi^{\pm}_n
\ ,\ B \Phi^{\pm}_n = b \Phi^{\pm}_n
\label{boson}
\ee
and the odd lowest weights: 
\be
\Psi^{-}_n\equiv\theta_{12}Z_{12}^{n}
\ ;\ \Psi^{+}_n\equiv\bar\theta_{12}Z_{12}^{n}
\ ;\ S \Psi^{\pm}_n = (n +\ell+\half)\Psi^{\pm}_n
\ ,\ B \Psi^{\pm}_n = (b\pm\half)\Psi^{\pm}_n
\label{fermion}
\ee
It is convenient to choose the chiral basis 
$D_1^{\pm}\Phi^{\pm}_n = 0$  for the even lowest weights.\\ 
Thus we have obtained the full set of lowest weights appearing 
in the expansion of $V_{\ell_1,b_1}\otimes V_{\ell_2,b_2}$ 
and this proves the direct sum decomposition 
in generic situation~(\ref{sum}).\\
The modifications in the case of the tensor product 
of chiral modules are evident:
the lowest weights in the tensor product of the chiral 
modules are obtained by imposing the chirality restrictions.
All lowest weights in the space 
$V_{\ell_1,\mp\ell_1}\otimes V_{\ell_2,b_2}~,~\ell_2\ne\pm b_2 $ 
are $\Phi^{\pm}_n$ and $\Psi^{\pm}_n$ and the 
lowest weights in the space 
$V_{\ell_1,\mp\ell_1}\otimes V_{\ell_2,\pm\ell_2}$ 
are $\Phi^{\pm}_n$.
All lowest weights in the space 
$V_{\ell_1,-\ell_1}\otimes V_{\ell_2,-\ell_2}$ 
are $\theta_{12}(z^{-}_{1}-z^{-}_{2})^n=\Psi^{-}_n$ and 
all lowest weights in the space 
$V_{\ell_1,\ell_1}\otimes V_{\ell_2,\ell_2}$ 
are $\bar\theta_{12}(z^{+}_{1}-z^{+}_{2})^n=\Psi^{+}_n$.

\section{Yang-Baxter equation and 
general operator $\R_{\vec\ell_1\vec\ell_2}(u)$}
\setcounter{equation}{0}

Let $V_{\ell_i,b_i}\ ;\ i=1,2,3$ be three lowest 
weight $s\ell(2|1)$-modules. 
We shall use the short-hand notation:
$$
\vec\ell=(\ell,b)\ ;\  V_{\vec\ell}=V_{\ell,b}
$$
Let us consider the three operators $\R_{\vec\ell_i\vec\ell_j}(u)$ 
which are acting in $V_{\vec\ell_i}\otimes V_{\vec\ell_j}$ and obey 
the Yang-Baxter equation in the space 
$V_{\vec\ell_1}\otimes V_{\vec\ell_2}\otimes V_{\vec\ell_3}$~\cite{KS}:
\be
\R_{\vec\ell_1\vec\ell_2}(u-v)\R_{\vec\ell_1\vec\ell_3}(u)
\R_{\vec\ell_2\vec\ell_3}(v)=
\R_{\vec\ell_2\vec\ell_3}(v)\R_{\vec\ell_1\vec\ell_3}(u)
\R_{\vec\ell_1\vec\ell_2}(u-v)
\label{YB}
\ee 
We are going to find the general solution 
$\R_{\vec\ell_i\vec\ell_j}(u)$ of Yang-Baxter 
equation by the following three steps.

First one obtains the operator $\R_{\vec f,\vec f}(u)$ 
in the simplest situation:
$\vec\ell_i = \vec f$ for all $i = 1,2,3$ so that the space 
$V_{\vec f}=V_{-\half,-\half}$ has 
the minimal possible dimension: $\dim V_{\vec f}=3$. 
In the second step we fix  $\vec\ell_i = \vec f $
for $i =1,2$ and obtain the solution 
$\R_{\vec f\vec\ell}$ for arbitrary $\vec\ell$.
In the third step we fix $\vec\ell_3 = \vec f $
and using the result for the operator 
$\R_{\vec f\vec\ell_1}$ we obtain and solve 
the defining equation for the 
general R-matrix $\R_{\vec\ell_i\vec\ell_j}(u)$.
It should be noted that the analogous approach was used for the 
derivation of the $s\ell(2)$-invariant R-matrix~\cite{KRS}.

\subsection{Fundamental solution $\R_{\vec f,\vec f}$}
\setcounter{equation}{0}

First one considers the simplest situation:
$\ell_i = -\half\ ,\ b_i = -\half \leftrightarrow \vec\ell_i = \vec f_i$.
We shall prove that the operator:
\be
\R_{\vec f_i\vec f_j}(u) = u + \eta P_{ij}
\ ;\ P_{ij}\equiv\sum_{AB} (-)^{\bar B}e^i_{AB}e^j_{BA}
\label{anz}
\ee 
where $e^i_{AB}$ are generators acting in 
the space $V_{\vec f_i}$, is the solution 
of the Yang-Baxter equation~\cite{KS}:
$$
\R_{\vec f_1\vec f_2}(u-v)\R_{\vec f_1\vec f_3}(u)
\R_{\vec f_2\vec f_3}(v)=
\R_{\vec f_2\vec f_3}(v)\R_{\vec f_1\vec f_3}(u)
\R_{\vec f_1\vec f_2}(u-v)
$$
Indeed, the proof is the following.
The Yang-Baxter equation has the simple form 
in short notations:  
$$
(u-v+\eta P_{12})(u+\eta P_{13} )(v+\eta P_{23}) = 
(v+\eta  P_{23})(u+\eta P_{13})(u-v+\eta P_{12})
$$
Comparing operator coefficients of $u^k$ on both 
sides of this equation yields:
\be
u^0:\  P_{12}P_{13}P_{23} = P_{23}P_{13}P_{12}
\label{1}
\ee
\be
u^1:\  P_{13}P_{23} + P_{12}P_{23} = 
P_{23}P_{12} + P_{23}P_{13}
\label{2}
\ee
Using~(\ref{ee}) one can prove that the 
operator $P_{ij}$ is the permutation:
$$
P_{ij}e^i_{AB} = e^j_{AB}P_{ij}
\Rightarrow P_{ij}P_{jk} = P_{ik}P_{ij} 
$$
and this commutation relation for $P_{ij}$ allows to check 
that eqs.~(\ref{1}),~(\ref{2}) hold and this proves that 
$\R_{\vec f\vec f}$~(\ref{anz}) obeys the Yang-Baxter equation.

\subsection{The solution for the 
operator $\R_{\vec f,\vec\ell}(u)$ }
\setcounter{equation}{0}

We fix  $\vec\ell_i = \vec f_i $
for $i = 1,2$ and obtain the solution 
$\R_{\vec f\vec\ell}$ for arbitrary $\vec\ell$.
The operator:
$$
\R_{\vec f\vec\ell}(u) = u  + 
\eta \sum_{AB} (-)^{\bar B}e_{AB}E_{BA},
$$ 
where $E_{AB}$ are generators in arbitrary 
representation $\vec\ell$, is the solution 
of the Yang-Baxter equation:
$$
\R_{\vec f_1\vec f_2}(u-v)\R_{\vec f_1\vec\ell_3}(u)
\R_{\vec f_2\vec\ell_3}(v)=
\R_{\vec f_2\vec\ell_3}(v)\R_{\vec f_1\vec\ell_3}(u)
\R_{\vec f_1\vec f_2}(u-v)
$$
The proof is the following~\cite{K} .
The Yang-Baxter equation has the simple form 
in short-hand notations:  
$$
(u-v+\eta P_{12})
(u+\eta e\otimes 1\otimes E)(v+\eta 1\otimes e\otimes E) = 
(v+\eta 1\otimes e\otimes E)(u+\eta e\otimes 1\otimes E)
(u-v+\eta P_{12})
$$
Matching operator coefficients of $u^k$ on both 
sides of this equality yields:
$$
u^0:\  P_{12}(e\otimes 1\otimes E) (1\otimes e\otimes E) = 
(1\otimes e\otimes E)( e\otimes 1\otimes E)P_{12}
$$
$$
u^1:\ (e\otimes 1\otimes E)(1\otimes e\otimes E) + 
P_{12}(1\otimes e\otimes E) = 
(1\otimes e\otimes E)P_{12} +
(1\otimes e\otimes E)(e\otimes 1\otimes E)
$$
The first equation is a simple consequence of the properties of $P_{12}$.
Using the fact that the matrices $e_{AB}$ 
form a basis we obtain that the second equation 
is equivalent to the following system of equations:
$$
E_{AB}\cdot E_{CD} - 
(-)^{(\bar{A}+\bar{B})(\bar{C}+\bar{D})}
E_{CD}\cdot E_{AB} = 
\delta_{CB} E_{AD} - 
(-)^{(\bar{A}+\bar{B})(\bar{C}+\bar{D})}
\delta_{AD} E_{CB}.
$$
This is nothing else but the commutation 
relations for generators $E_{AB}$.

Let us represent the operator under consideration:
$$
\R_{\vec f\vec\ell}(u) = u  + 
\eta \sum_{AB} (-)^{\bar B}e_{AB}E_{BA}
$$
in the matrix form in the standard basis 
with the grading $\bar 1,\bar 3 = 0\ ;\  \bar 2 = 1$:
$$
e_1 = 
\left (\begin{array}{c}
1 \\ 0 \\0 \\   
\end{array} \right )
\ ;\ e_2 =  
\left (\begin{array}{ccc}
0 \\1  \\ 0    
\end{array} \right )
\ ;\ e_3 = 
\left (\begin{array}{ccc}
0 \\0  \\ 1    
\end{array} \right )
$$ 
We shall use the following definition of 
the matrix of an operator:
\be
F e_A = e_B F_{BA},
\label{matr}
\ee
which leads to the validity of the common rule for the matrix product, 
i.e. without additional sign factors,
$$
FG e_A = e_B (FG)_{BA}\ ;\  (FG)_{BA} = F_{BC}G_{CA}.
$$
Let us calculate the matrix of R-operator using 
the definitions~(\ref{matr}) and~(\ref{ee}):
$$
(e_{AB}E_{BA})e_C = 
(-)^{(\bar B+\bar A)\bar C}e_{AB}e_C E_{BA} =
(-)^{(\bar B+\bar A)\bar C}e_D(e_{AB})_{DC}E_{BA}.
$$  
Therefore the matrix element of operator $e_{AB}E_{BA}$ 
has the form
$$
\sum_{AB}(-)^{\bar B}(e_{AB}E_{BA})_{CD} = 
(-)^{\bar D\bar C}E_{DC}. 
$$
Finally one obtains
$$
\R_{\vec f\vec\ell}(u) = u  + 
\eta \sum_{AB} (-)^{\bar B}e_{AB}E_{BA} =
\left (\begin{array}{ccc}
u+\eta E_{11} & \eta E_{21} & \eta E_{31} \\
\eta E_{12} & u-\eta E_{22}  & \eta E_{32} \\
\eta E_{13} & \eta E_{23}  &  u+\eta E_{33}
\end{array} \right ).
$$
This is the expression for the $g\ell(2|1)$-invariant 
R-matrix. The $s\ell(2|1)$-invariant 
R-matrix can be derived from this result in 
a simple way:
the operator $K_1 = E_{11}+E_{22}+E_{33}$ 
belongs to the center of the algebra and therefore 
the R-operator $\R_{\vec f\vec\ell}(u-\eta K_1)$ is 
also a solution of the Yang-Baxter equation.
Using the connection between $E_{AB}$ and 
$s\ell(2|1)$-generators we obtain the 
$s\ell(2|1)$-invariant R-matrix~\cite{FPT}
$$
\R_{\vec f\vec\ell}(u-\eta K_1)  =
\left (\begin{array}{ccc}
u+\eta (S+B) & -\eta W^{-} & \eta S^{-} \\
\eta V^{+} & u+2\eta B  & \eta V^{-} \\
\eta S^{+} & \eta W^{+}  &  u+\eta(B-S) 
\end{array} \right ).
$$

\subsection{General R-matrix $\R_{\vec\ell_1\vec\ell_2}(u)$}
\setcounter{equation}{0}

To obtain the defining relation for the general 
$\R$-operator we consider the special case 
${\vec\ell}_3=\vec f$ in~(\ref{YB}).  
Then one can choose the above matrix realization 
in $V_{\vec\ell_3}$ and the operators 
$\R_{\vec\ell_1\vec f}$,$\R_{\vec\ell_2\vec f}$ 
are linear functions of spectral parameter $u$ 
in this particular case
$$
\R_{\vec\ell_i\vec f}(u-\eta K_1) = u+\eta \R_{i}
\ ;\ \R_{i}=
\left (\begin{array}{ccc}
S_i+B_i & -W^{-}_i & S^{-}_i \\
V^{+}_i & 2 B_i  & V^{-}_i \\
S^{+}_i & W^{+}_i & B_i-S_i
\end{array} \right )
\ ;\ i=1,2
$$
Now the general R-matrix $\R_{\vec\ell_1\vec\ell_2}(u)$ 
acting in the tensor product $V_{\vec\ell_1}\otimes V_{\vec\ell_2}$
of arbitrary modules, is fixed by the condition
\be
\R_{\vec\ell_1\vec\ell_2}(u-v)
\R_{\vec\ell_1\vec f}(u-\eta K_1)\R_{\vec\ell_2\vec f}(v-\eta K_1) = 
\R_{\vec\ell_2\vec f}(v-\eta K_1)\R_{\vec\ell_1\vec f}(u-\eta K_1)
\R_{\vec\ell_1\vec\ell_2}(u-v)
\label{m}
\ee
or equivalently:
$$
\R_{\vec\ell_1\vec\ell_2}(u-v)\left(
\frac{uv}{\eta^2} + \frac{u+v}{2\eta}(\R_1+ \R_2)+ 
\frac{u-v}{2\eta}(\R_2-\R_1)+\R_1\R_2\right) =
$$
$$
=\left(\frac{uv}{\eta^2} + \frac{v+u}{2\eta}(\R_2+ \R_1)+ 
\frac{v-u}{2\eta}(\R_1-\R_2)+\R_2\R_1\right)\R_{\vec\ell_1\vec\ell_2}(u-v).
$$
After separation of $u+v$ and $u-v$ dependence we 
obtain two equations($u-v\to u$):
\be
\R_{\vec\ell_1\vec\ell_2}(u)(\R_1+ \R_2)=
(\R_1+ \R_2)\R_{\vec\ell_1\vec\ell_2}(u)
\label{invar}
\ee
\be
\R_{\vec\ell_1\vec\ell_2}(u)\left(\frac{u}{2\eta}(\R_2-\R_1)+\R_1\R_2\right) =
\left(\frac{u}{2\eta}(\R_2-\R_1)+\R_2\R_1\right)\R_{\vec\ell_1\vec\ell_2}(u)
\label{info}
\ee
The first equation~(\ref{invar}) expresses the 
fact that $\R(u)$ has to be invariant 
with respect to the action of $s\ell(2|1)$-algebra and the 
second equation is the wanted defining relation for the 
operator $\R_{\vec\ell_1\vec\ell_2}(u)$.\\ 
The $s\ell(2|1)$-invariance of the operator 
$\R_{\vec\ell_1\vec\ell_2}(u)$ allows to simplify the problem.
Due to $s\ell(2|1)$-invariance 
any eigenspace of the operator $\R_{\vec\ell_1\vec\ell_2}$
is a lowest weight $s\ell(2|1)$-module 
generated by some lowest weight eigenvector. 
Therefore without loss of generality we can 
solve the defining relation~(\ref{info}) in the space of lowest weights.
Let us consider in more details the structure of 
eigenspace of the $s\ell(2|1)$-invariant operator 
acting on the tensor 
product~$V_{\ell_1,b_1}\otimes V_{\ell_2,b_2}$.
As we have seen from direct sum decomposition: 
$$
V_{\ell_1,b_1}\otimes V_{\ell_2,b_2} =  V_{\ell,b}+
2\sum_{n=1}^{\infty} V_{\ell+n,b}+
\sum_{n=0}^{\infty} V_{\ell+n+\half,b-\half}+
\sum_{n=0}^{\infty} V_{\ell+n+\half,b+\half}
$$
for every fixed $n$ the space of lowest weight vectors with 
eigenvalue $b$ is two-dimensional and the ones with
eigenvalues $b\pm \half$ are one-dimensional.
Therefore the operator $\R_{\vec\ell_1\vec\ell_2}$
is diagonal on odd lowest weight vectors 
$\Psi^{+}_{n}$ and $\Psi^{-}_{n}$
but acts non-trivially on the two-dimensional subspace of 
even lowest weight vectors spanned on 
$\Phi^{+}_{n}$ and $\Phi^{-}_{n}$.\\
In matrix form we have:
\be
\R_{\vec\ell_1\vec\ell_2}(u)
\left (\begin{array}{c}
\Phi^{+}_{n} \\  \Phi^{-}_{n} \\ \Psi^{+}_{n}\\ 
\Psi^{-}_{n} 
\end{array} \right ) =
\left (\begin{array}{cccc}
A_{n}(u) & B_{n}(u) & 0 & 0 \\
C_{n}(u) & D_{n}(u) & 0 & 0 \\
0 & 0 & F_{n}(u) & 0 \\
0 & 0 & 0 & E_{n}(u) \\
\end{array} \right )
\left (\begin{array}{c}
\Phi^{+}_{n} \\  \Phi^{-}_{n} \\ \Psi^{+}_{n}\\ 
\Psi^{-}_{n} 
\end{array} \right )
\label{ABCD}
\ee
The matrix relation~(\ref{info}) leads to a set of 
recurrence relations for the coefficients $A_{n},...,E_{n}$.
Some details of calculations can be found in Appendix and 
here we present the final expression for the 
general solution of these recurrence relations:
\be
A_{n}(u)=(-1)^{n+1}
\frac{\Gamma\left(\u+\ell_n+1\right)}
{\Gamma\left(-\u+\ell_n\right)}
\cdot\frac{\u+b_{1}-b_{2}}{(\ell_1-b_1)(\ell_2+b_2)}
\label{solABCD}
\ee
$$
B_{n}(u)=(-1)^{n}
\frac{\Gamma\left(\u+\ell_n+1\right)}
{\Gamma\left(-\u+\ell_n+1\right)}
\cdot\frac{(\ell_1+b_1)(\ell_2-b_2)}{(\ell_1-b_1)(\ell_2+b_2)}
$$
$$
C_{n}(u)=(-1)^{n}
\frac{\Gamma\left(\u+\ell_n\right)}
{\Gamma\left(-\u+\ell_n\right)}
\ ;\  D_{n}(u)=(-1)^{n+1}
\frac{\Gamma\left(\u+\ell_n\right)}
{\Gamma\left(-\u+\ell_n+1\right)}
\cdot
$$
$$
\cdot\frac{
(\ell_2-b_2)(\ell_2+b_2)\left(\u-b_{1}-b_{2}\right)-
\left(\u+b_{1}+b_{2}\right)
\left(\u-b_2-\ell_1\right)\left(\u-b_2+\ell_1\right)}
{(\ell_1-b_1)(\ell_2+b_2)}
$$
$$
E_{n}(u)=(-1)^{n}
\frac{\Gamma\left(\u+\ell_n+1\right)}
{\Gamma\left(-\u+\ell_n+1\right)}
\cdot\frac{(\u+b_1-\ell_2)(\u+b_1+\ell_2)}
{(\ell_1-b_1)(\ell_2+b_2)}
$$
$$
F_{n}(u)=(-1)^{n}
\frac{\Gamma\left(\u+\ell_n+1\right)}
{\Gamma\left(-\u+\ell_n+1\right)}
\cdot\frac{(\u-b_2-\ell_1)(\u-b_2+\ell_1)}{(\ell_1-b_1)(\ell_2+b_2)}
$$
where we used the notations:
$$
\ell_n\equiv n + \ell_1+\ell_2
\ ;\ \u\equiv \frac{u}{\eta}+ b_1-b_2.
$$
As usual the obtained general solution of the Yang-Baxter 
equation is fixed up to overall normalization.
We choose the normalization such that the R-matrix 
coincides with the permutation operator for 
$u=0$ and $\vec\ell_1=\vec\ell_2$.\\
The obtained R-matrix~(\ref{ABCD}) acts on the 
space of two-point functions which are polynomials in 
$z_i,\theta_i,\bar\theta_i,~i=1,2$.
This holds also if the representation parameters 
$\ell_i,b_i,~i=1,2$ correspond to chiral or antichiral cases.
If one or both modules in the tensor product are 
chiral or antichiral then the tensor product
representation space is a proper subspace of the 
space of all two-point polynomials~(compare~(\ref{sum})). 
It is important to observe that in these cases the 
action of R-matrix can be consistently restricted 
to the corresponding subspace.
Indeed, after multiplying with the overall factor
$(\ell_1-b_1)(\ell_2+b_2)$, the matrix becomes 
triangular in these cases in the way as expected.
We list the reduced R-matrices in all special cases 
involving chiral representations in Appendix B.

\section{Homogeneous periodic chain}
\setcounter{equation}{0}

\subsection{Commuting transfer matrices}
\setcounter{equation}{0}

Let us construct the set of commuting $s\ell(2|1)$-invariant
operators the generating function of which is the 
transfer-matrix $\T_{\vec m}(u)$.  
We construct $\T_{\vec m}(u)$ as the supertrace
of a monodromy matrix built 
of the elementary {\R}-matrix blocks~\cite{K}.

We introduce the $N$ spaces $V_{\vec\ell_i}$ 
and $N$ operators $\R_{\vec m,\vec\ell_i}(u)$:
$$
\R_{\vec m,\vec\ell_i}(u): 
V_{\vec m}\otimes V_{\vec\ell_i} 
\mapsto V_{\vec m}\otimes V_{\vec\ell_i} 
$$
The periodicity convention $N+1\equiv 1$ is implied. 
The monodromy matrix
\be
\R_{\vec m}(u)\equiv \R_{\vec m,\vec\ell_1}(u-c_1)\ldots
\R_{\vec m,\vec\ell_N}(u-c_N)
\label{monodr}
\ee
acts then on the space 
$V_{\vec m}\otimes V_{\vec\ell_1}\otimes \ldots\otimes V_{\vec\ell_N}$,
and $\T_{\vec m}(u)$ is obtained by taking the supertrace in 
the auxiliary space $V_{\vec m}$:
$$
\T_{\vec m}(u)=\str_{V_{\vec m}}\R_{\vec m}(u).
$$
These monodromy matrices form the commutative family:
\be
 \T_{{\vec m}_1}(u)\T_{{\vec m}_2}(v) = 
\T_{{\vec m}_2}(v)\T_{{\vec m}_1}(u)
\label{comm}
\ee
The relation \eref{comm} follows from the fact that there exists the 
operator $\R_{{\vec m}_1,{\vec m}_2}$ such that
$$
\R_{{\vec m}_1,{\vec m}_2}(u-v)\R_{{\vec m}_1,\vec\ell_i}(u)
\R_{\vec m_2,\vec\ell_i}(v)=
\R_{\vec m_2,\vec\ell_i}(v)\R_{\vec m_1,\vec\ell_i}(u)
\R_{\vec m_1,\vec m_2}(u-v)
$$
The $\R$-operator is even. Therefore by using standard arguments 
one derives an analogous equation for the monodromy matrices:
\be
\R_{{\vec m}_1,{\vec m}_2}(u-v)\R_{{\vec m}_1}(u)
\R_{\vec m_2}(v)=
\R_{\vec m_2}(v)\R_{\vec m_1}(u)
\R_{\vec m_1,\vec m_2}(u-v)
\label{cm}
\ee
From this one can derive easily that corresponding 
traces and supertraces are commuting operators separately:
$$
\T^{+}_{\vec m}(u)=\tr_{V_{\vec m}}\R_{\vec m}(u)
\ ;\ \T^{+}_{{\vec m}_1}(u)\T^{+}_{{\vec m}_2}(v) = 
\T^{+}_{{\vec m}_2}(v)\T^{+}_{{\vec m}_1}(u)
$$
$$
\T^{-}_{\vec m}(u)=\str_{V_{\vec m}}\R_{\vec m}(u)
\ ;\ \T^{-}_{{\vec m}_1}(u)\T^{-}_{{\vec m}_2}(v) = 
\T^{-}_{{\vec m}_2}(v)\T^{-}_{{\vec m}_1}(u)
$$
but only $\T_{\vec m}(u)\equiv\T^{-}_{\vec m}(u)$ 
is the generating function for the 
$s\ell(2|1)$-invariant operators. 

Instead of giving the general proof we demonstrate all this on the 
example of $\T_{\vec f}(u)$, where the auxiliary space corresponds to the 
fundamental representation.
Let us represent $\T_{\vec f}(u)$ in the form:
$$
\T_{\vec m}(u) = e_{AB}\T_{AB}(u)  
$$
where operators $\T_{AB}(u)$  act in tensor product
$V_{\vec\ell_1}\otimes\cdots\otimes V_{\vec\ell_n}$ 
and we assume the summation over repeated indices.
The general equation~(\ref{cm}) has the form in this case
$$
\left[u-v+\eta(-)^{\bar G}e^i_{FG}e^j_{GF}\right]
e^i_{AB}\T_{AB}(u)~ e^j_{CD}\T_{CD}(v)=
$$
$$
= e^j_{CD}\T_{CD}(v)~ e^i_{AB}\T_{AB}(u)
\left[u-v+\eta(-)^{\bar G}e^i_{FG}e^j_{GF}\right].
$$
The traces and supertraces of generators $e_{CD}$ 
are calculated as follows:
$$
\tr e_{AB}\equiv (e_{AB})_{CC}=\delta_{AB}
\ ;\ \str e_{AB}\equiv (-)^{\bar C}(e_{AB})_{CC}
=(-)^{\bar A}\delta_{AB}
$$ 
Using $e_{AB}e_{CD}=\delta_{CB}e_{AD}$ and taking 
(super-)traces in corresponding spaces one easily obtains:
$$
\tr: \T_{AA}(u) \T_{CC}(v)
= \T_{CC}(v) \T_{AA}(u),
$$ 
$$
\str: (-)^{\bar A}\T_{AA}(u) (-)^{\bar C}\T_{CC}(v)
= (-)^{\bar C}\T_{CC}(v) (-)^{\bar A}\T_{AA}(u),
$$
The $g\ell(2|1)$-invariance can be demonstrated in the simplest 
example $N=2$. The generalization to 
arbitrary $n$ is straightforward.  
$$
\T^{+}_{\vec f}(u)=\tr 
\left(u+\eta(-)^{\bar B}e_{AB}E^1_{BA}\right)
\left(u+\eta(-)^{\bar D}e_{CD}E^2_{DC}\right)=
$$
$$
=3 u^2 + \eta u (-)^{\bar A}\left(E^1_{AA}+E^2_{AA}\right)
+ \eta^2 E^1_{AB}E^2_{BA}
$$
$$
\T^{-}_{\vec f}(u)=\str 
\left(u+\eta(-)^{\bar B}e_{AB}E^1_{BA}\right)
\left(u+\eta(-)^{\bar D}e_{CD}E^2_{DC}\right)=
$$
$$
=u^2 + \eta u \left(E^1_{AA}+E^2_{AA}\right)
 + \eta^2 (-)^{\bar B} E^1_{AB}E^2_{BA}.
$$
We have seen~((\ref{K}) and discussion before) that only operators 
entering~$\T^{-}_{\vec f}(u)$ are $g\ell(2|1)$-invariant.

\subsection{Local Hamiltonians}
\setcounter{equation}{0}

If one fixes the arbitrary representation $\vec m =\vec\ell$ and
the same representations 
$\vec \ell_1=\vec \ell_2 =...=\vec \ell_N=\vec\ell$ in
remaining spaces $V_{\vec\ell_i}$ we obtain the 
generating functions of Hamiltonians.
The local Hamiltonians can be obtained for the homogeneous
($c_i=0$) chain.The whole construction is quite general~\cite{H}.
Let us calculate the first two coefficients in the Taylor expansion
of the operator
$$
\T_{\vec\ell}(u)=\str_{V_{\vec\ell}}\R_{\vec\ell,\vec\ell_1}(u)
\ldots\R_{\vec\ell,\vec\ell_N}(u)
$$
around the point $u=0$.
It is easy to see from the derived expression for the 
R-matrix that by condition $\vec\ell_i=\vec\ell_j=\vec\ell$ 
the point $u = 0$ is regular: 
\be
\R_{\vec\ell_i,\vec\ell_j}(0) = \P_{\vec\ell_i,\vec\ell_j}
\label{reg}
\ee
where $\P$ is permutation operator.\\ 
Indeed, the permutation operator $\P$ acts as follows 
on the lowest weight basis:
$$
\P
\left (\begin{array}{c}
\Phi^{+}_{n} \\  \Phi^{-}_{n} \\ \Psi^{+}_{n}\\ 
\Psi^{-}_{n} 
\end{array} \right ) = (-1)^n
\left (\begin{array}{cccc}
0 & 1 & 0 & 0 \\
1 & 0 & 0 & 0 \\
0 & 0 & -1 & 0 \\
0 & 0 & 0 & -1 \\
\end{array} \right )
\left (\begin{array}{c}
\Phi^{+}_{n} \\  \Phi^{-}_{n} \\ \Psi^{+}_{n}\\ 
\Psi^{-}_{n} 
\end{array} \right )
$$ 
The expression for matrix coefficients of 
R-operator takes the simple form in homogeneous 
case~($\ell_1=\ell_2,\  b_1=b_2$):
\be
A_{n}(u)=(-1)^{n+1}
\frac{\Gamma\left(\u+\ell_n+1\right)}
{\Gamma\left(-\u+\ell_n\right)}
\cdot\frac{\u}{(\ell+b)(\ell-b)}
\label{solABCDhom}
\ee
$$
B_{n}(u)=(-1)^{n}
\frac{\Gamma\left(\u+\ell_n+1\right)}
{\Gamma\left(-\u+\ell_n+1\right)}
\ ;\  C_{n}(u)=(-1)^{n}
\frac{\Gamma\left(\u+\ell_n\right)}
{\Gamma\left(-\u+\ell_n\right)}
$$
$$
D_{n}(u)=(-1)^{n+1}
\frac{\Gamma\left(\u+\ell_n\right)}
{\Gamma\left(-\u+\ell_n+1\right)}
\cdot \frac{\u\left(\u^2-2\ell^2-2b^2\right)}{(\ell+b)(\ell-b)}
$$
$$
E_{n}(u)=(-1)^{n}
\frac{\Gamma\left(\u+\ell_n+1\right)}
{\Gamma\left(-\u+\ell_n+1\right)}
\cdot\frac{(\u+b-\ell)(\u+b+\ell)}{(\ell+b)(\ell-b)}
$$
$$
F_{n}(u)=(-1)^{n}
\frac{\Gamma\left(\u+\ell_n+1\right)}
{\Gamma\left(-\u+\ell_n+1\right)}
\cdot\frac{(\u-b-\ell)(\u-b+\ell)}{(\ell+b)(\ell-b)},
$$
where 
$$
\ell_1=\ell_2=\ell\ ;\  b_1=b_2=b
\ ;\ \u\equiv \frac{u}{\eta}
\ ;\ \ell_n \equiv n + 2\ell.
$$
The equality~(\ref{reg}) can be easily checked.\\
The first coefficient in the Taylor expansion
of the operator $\T_{\vec\ell}(u)$
is proportional to the operator of cyclic shift:
$$
\T_{\vec\ell}(0) =\str_{V_{\vec\ell}}
\P_{\vec\ell,\vec\ell_1}\P_{\vec\ell,\vec\ell_2}\cdots 
\P_{\vec\ell,\vec\ell_N}
= const\cdot \P_{\vec\ell_N,\vec\ell_{N-1}}
\P_{\vec\ell_{N-1},\vec\ell_{N-2}}\cdots \P_{\vec\ell_2,\vec\ell_1}
$$
This is readily checked.
First we move $\P_{\vec\ell,\vec\ell_1}$ to the right, then 
$\P_{\vec\ell_1,\vec\ell_2}$ to the right an so on and 
after all we use $\str_{V_{\vec\ell}}\P_{\vec\ell,\vec\ell_1} = const$.

The second coefficient has the form:
$$
\T^{\prime}_{\vec\ell}(0)=
\sum_i \str_{V_{\vec\ell}}
\P_{\vec\ell,\vec\ell_1}\cdots
\R^{\prime}_{\vec\ell,\vec\ell_i}(0)\cdots \P_{\vec\ell,\vec\ell_N}
$$
In order to simplify this expression let us consider the i-th 
term and move the $\R^{\prime}_{\vec\ell,\vec\ell_i}(0)$ 
to the right:
$$
\str_{V_{\vec\ell}}
\P_{\vec\ell,\vec\ell_1}\cdots 
\P_{\vec\ell,\vec\ell_{i-1}}
\R^{\prime}_{\vec\ell,\vec\ell_i}(0)
\P_{\vec\ell,\vec\ell_{i+1}}\cdots 
\P_{\vec\ell,\vec\ell_n} =
\str_{V_{\vec\ell}}
\P_{\vec\ell,\vec\ell_1}\cdots 
\P_{\vec\ell,\vec\ell_{i-1}}
\P_{\vec\ell,\vec\ell_{i+1}}\cdots 
\P_{\vec\ell,\vec\ell_n}
\R^{\prime}_{\vec\ell_{i+1},\vec\ell_i}(0)
$$
By moving first $\P_{\vec\ell,\vec\ell_1}$ to the right, 
then $\P_{\vec\ell_1,\vec\ell_2}$ and so on we transform 
the remaining term 
$$
\str_{V_{\vec\ell}}
\P_{\vec\ell,\vec\ell_1}\cdots 
\P_{\vec\ell,\vec\ell_{i-1}}
\P_{\vec\ell,\vec\ell_{i+1}}\cdots 
\P_{\vec\ell,\vec\ell_N} = const\cdot
\P_{\vec\ell_{N-1},\vec\ell_N}\cdots 
\P_{\vec\ell_{i-1},\vec\ell_{i+1}}\cdots 
\P_{\vec\ell_1,\vec\ell_2}
$$
On the last stage we multiply the obtained 
expression by the operator 
$$
\T^{-1}_{\vec\ell}(0)= (const)^{-1}\P_{\vec\ell_1,\vec\ell_{2}}
\cdots \P_{\vec\ell_{N-1},\vec\ell_N}
$$ 
from the left and obtain the following expression:
\begin{equation}
\T^{-1}_{\vec\ell}(0)\T^{\prime}_{\vec\ell}(0) =
\sum_i \P_{\vec\ell_{i+1},\vec\ell_{i}}\cdot
\R^{\prime}_{\vec\ell_{i+1},\vec\ell_{i}}(0)= 
\sum_i \R^{\prime}_{\vec\ell_{i},\vec\ell_{i+1}}(0)
\cdot\P_{\vec\ell_{i},\vec\ell_{i+1}}=
\sum_i \H_{\vec\ell_{i},\vec\ell_{i+1}}.
\label{hamilt}
\end{equation}
The resulting operator can be chosen as the Hamiltonian.
It commutes with the integrals of motions generated 
by~$\T_{\vec\ell}(u)$ and is a sum of operators 
acting on two adjacent sites only. 
The two-particle Hamiltonians in the sum are  
$$
\H_{\vec\ell_{i},\vec\ell_{i+1}} = 
\R^{\prime}_{\vec\ell_{i},\vec\ell_{i+1}}(0)
\cdot\P_{\vec\ell_{i},\vec\ell_{i+1}}
$$
and have the following matrix elements: 
\be
\H_{\vec\ell_{i},\vec\ell_{i+1}} = \eta^{-1}
\left (\begin{array}{cccc}
2\psi(\ell_n+1) & 
-\frac{\ell_n}{(\ell-b)(\ell+b)} & 0 & 0 \\
-\frac{2}{\ell_n}\frac{\ell^2+b^2}{(\ell-b)(\ell+b)} & 
2\psi(\ell_n) & 0 & 0 \\
0 & 0 & 2\psi(\ell_n+1)+\frac{2b}{(\ell-b)(\ell+b)} & 0 \\
0 & 0 & 0 & 2\psi(\ell_n+1)-\frac{2b}{(\ell-b)(\ell+b)} \\
\end{array} \right ).
\label{Hper}
\ee
The eigenvalues of this matrix and corresponding eigenvalues 
can be easily calculated.

\section{Inhomogeneous open chain}
\setcounter{equation}{0}

\subsection{Integrals of motion}
\setcounter{equation}{0}

In this section we shall consider the open spin chain~\cite{S}.
To start with we introduce the operator 
${\cal R}_{\vec\ell_{1},\vec\ell_{2}}(u)$:
$$
{\cal R}_{\vec\ell_{1},\vec\ell_{2}}(u)\equiv
\R_{\vec\ell_{1},\vec\ell_{2}}(u)\cdot
\R^{-1}_{\vec\ell_{1},\vec\ell_{2}}(-u).
$$
It is possible to derive the following commutation relation
for ${\cal R}(u)$:
$$
\R_{\vec m_1,\vec m_2}(u-v)\cdot{\cal R}_{\vec m_1,\vec\ell_i}(u)\cdot
\R_{\vec m_1,\vec m_2}(u+v)\cdot {\cal R}_{\vec m_2,\vec\ell_i}(v)=
{\cal R}_{\vec m_2,\vec\ell_i}(v)\cdot \R_{\vec m_1,\vec m_2}(u+v)\cdot
{\cal R}_{\vec m_1,\vec\ell_i}(u)\cdot \R_{\vec m_1,\vec m_2}(u-v).
$$
It is evident that the operator ${\cal R}_{\vec m}(u)$ 
constructed from the monodromy matrix~(\ref{monodr})
$$
{\cal R}_{\vec m}(u)\equiv
\R_{\vec m}(u)\cdot \R^{-1}_{\vec m}(-u),
$$
satisfies the analogous equation:
$$
\R_{\vec m_1,\vec m_2}(u-v)\cdot{\cal R}_{\vec m_1}(u)\cdot
\R_{\vec m_1,\vec m_2}(u+v)\cdot {\cal R}_{\vec m_2}(v)=
{\cal R}_{\vec m_2}(v)\cdot \R_{\vec m_1,\vec m_2}(u+v)\cdot
{\cal R}_{\vec m_1}(u)\cdot \R_{\vec m_1,\vec m_2}(u-v).
$$
Using this equation it is possible to show that 
corresponding supertraces are commuting operators:
$$
\TT_{{\vec m}_1}(u)\TT_{{\vec m}_2}(v) = 
\TT_{{\vec m}_2}(v)\TT_{{\vec m}_1}(u)
\ ;\  \TT_{\vec m}(u) = 
\str_{V_{\vec m}}{\cal R}_{\vec m}(u)=
\str_{V_{\vec m}}\R_{\vec m}(u)\cdot \R^{-1}_{\vec m}(-u).
$$
If one fixes the representation $\vec m = \vec f$ we 
obtain the generating function of 
integrals of motions for the open chain.

\subsection{Local Hamiltonian}
\setcounter{equation}{0}

Let us suppose that representations $\vec\ell_i$ and shifts $c_i$ in
the product:
$$
\R_{\vec\ell}(u)\equiv
\R_{\vec\ell,\vec\ell_1}(u + c_1)\R_{\vec\ell,\vec\ell_2}(u+c_2)...
\R_{\vec\ell,\vec\ell_N}(u+c_N)
$$
are fixed as follows:
$$
\vec\ell_2=\vec\ell_3=...\vec\ell_{N-1}=\vec\ell
\ ;\ c_2=c_3=...c_{N-1}=0
\Rightarrow  \R_{\vec\ell_i,\vec\ell}(0)=
\P_{\vec\ell_i,\vec\ell}~,~i=2,...,n-1,
$$
where $\P$ is simply the permutation.
In the case of~$s\ell(2)$ there exist integrable nearest neighbour    
interactions for this slightly inhomogeneous chain~\cite{DKM}.
We extend this result to the case of~$s\ell(2|1)$ and 
construct the corresponding Hamiltonians.\\    
The R-matrix~(\ref{ABCD}) obeys the equation:
\be
\R_{\vec\ell_1,\vec\ell_2}(u)\cdot 
\R_{\vec\ell_2,\vec\ell_1}(- u) = P(u)\cdot 1
\label{unitar}
\ee
where right side of equation is proportional of unit operator and 
$P(u)$ is the function:
$$
P(u)\equiv\frac{(\u+b_1-\ell_2)(\u+b_1+\ell_2)(\u-b_2+\ell_1)(\u-b_2-\ell_1)}
{(\ell_1-b_1)(\ell_1+b_1)(\ell_2-b_2)(\ell_2+b_2)},
$$
which follows from the expression for matrix elements 
of the R-matrix~(\ref{solABCD}).
The explicit form of the matrix 
$\R_{\vec\ell_2,\vec\ell_1}(- u)$ is the following:
$$
\R_{\vec\ell_2\vec\ell_1}(-u) =
\left (\begin{array}{cccc}
\bar{D}_{n}(-u) & \bar{C}_{n}(-u) & 0 & 0 \\
\bar{B}_{n}(-u) & \bar{A}_{n}(-u) & 0 & 0 \\
0 & 0 & \bar{F}_{n}(-u) & 0 \\
0 & 0 & 0 &\bar{E}_{-n}(u) \\
\end{array} \right )
$$
where the matrix elements~$\bar{A}_n,\bar{B}_n,...$ are obtained 
from the elements~$A_n,B_n,...$ by formal change of variables
$\ell_1\leftrightarrow\ell_2$ and $b_1\leftrightarrow b_2$. 
Due to equation~(\ref{unitar}) we have
\be
\R^{-1}_{\vec\ell_1,\vec\ell_2}(u)\sim 
\R_{\vec\ell_2,\vec\ell_1}(- u)
\label{invers}
\ee
so that we can use the operator 
$\R_{\vec\ell_2,\vec\ell_1}(-u)$ instead of operator 
$\R^{-1}_{\vec\ell_1,\vec\ell_2}(u)$.
This changes the normalization of the 
operator~$\TT_{\vec m}(u)$ only.\\ 
Let us calculate the first two coefficients in the Taylor expansion
of the operator
$$
\TT_{\vec\ell}(u) = \str_{V_{\vec\ell}}
\R_{\vec\ell,\vec\ell_1}(u+c_1)
\ldots\R_{\vec\ell,\vec\ell_N}(u+c_N)
\R_{\vec\ell_N,\vec\ell}(u-c_N)
\ldots\R_{\vec\ell_1,\vec\ell}(u-c_1).
$$
The first coefficient is proportional to the unit operator for arbitrary
representations $\vec\ell_i$ and 
shifts $c_i$ due to the property~(\ref{invers}),
$$
\TT_{\vec\ell}(0) = \str_{V_{\vec\ell}}
\R_{\vec\ell,\vec\ell_1}(c_1)
\ldots\R_{\vec\ell,\vec\ell_N}(c_N)
\R_{\vec\ell_N,\vec\ell}(-c_N)
\ldots\R_{\vec\ell_1,\vec\ell}(-c_1) 
\sim \str_{V_{\vec\ell}} 1.
$$
Let us introduce the short-hand notation
\be
\H_{\vec\ell,\vec m}(c) = 
\R^{\prime}_{\vec\ell,\vec m}(c)\R_{\vec m,\vec\ell}(-c)+
\R_{\vec\ell,\vec m}(c)\R^{\prime}_{\vec m,\vec\ell}(-c)
\label{Hop}
\ee
and calculate the expression for 
$\TT^{\prime}_{\vec\ell}(0)$ which contains several terms ,
$$
\TT^{\prime}_{\vec\ell}(0)=
\str_{V_{\vec\ell}}\H_{\vec\ell,\vec\ell_1}(c_1)+
\str_{V_{\vec\ell}}\R_{\vec\ell,\vec\ell_1}(c_1)
\H_{\vec\ell,\vec\ell_2}(0)\R_{\vec\ell_1,\vec\ell}(-c_1)+
$$
$$
+\sum_{i=2...N} 
\str_{V_{\vec\ell}}\R_{\vec\ell,\vec\ell_1}(c_1)\cdots
\P_{\vec\ell,\vec\ell_{i-1}}\H_{\vec\ell,\vec\ell_i}(c_i)
\P_{\vec\ell_{i-1},\vec\ell}\cdots\R_{\vec\ell_1,\vec\ell}(-c_1).
$$
Note that this formula is true for $c_2 =\cdots c_{N-1}= 0$ only.
Let us consider each term separately. 
The first term is constant
$$
\str_{V_{\vec\ell}}\H_{\vec\ell,\vec\ell_i}(c_i) = const
$$
The $i$-th term in the sum can be transformed easily 
to the simpler expression
$$
\tr_{V_{\vec\ell}}\R_{\vec\ell,\vec\ell_1}(c_1)\cdots
\P_{\vec\ell,\vec\ell_{i-1}}\H_{\vec\ell,\vec\ell_i}(c_i)
\P_{\vec\ell_{i-1},\vec\ell}\cdots\R_{\vec\ell_1,\vec\ell}(-c_1)=
\H_{\vec\ell_{i-1},\vec\ell_i}(c_i)\cdot \str_{V_{\vec\ell}} 1
$$
It turns out that the second term can be transformed to the 
analogous form too (remember $c_2 = 0$)
\be
\str_{V_{\vec\ell}}\R_{\vec\ell,\vec\ell_1}(c_1)
\H_{\vec\ell,\vec\ell_2}(0)\R_{\vec\ell_1,\vec\ell}(-c_1) =
\H_{\vec\ell_1,\vec\ell_2}(-c_1)\cdot \str_{V_{\vec\ell}} 1 + const
\label{h}
\ee
For the proof we start from the Yang-Baxter equation
$$
\R_{{\vec\ell}_2,{\vec\ell}_1}(u)\R_{{\vec\ell}_2,\vec\ell}(u+v)
\R_{\vec\ell_1,\vec\ell}(v)=
\R_{\vec\ell_1,\vec\ell}(v)\R_{\vec\ell_2,\vec\ell}(u+v)
\R_{\vec\ell_2,\vec\ell_1}(u),
$$
differentiate this equation with 
respect to $u$ and then put $v=-u$:
$$
\R^{\prime}_{{\vec\ell}_2,{\vec\ell}_1}(u)\P_{{\vec\ell}_2,\vec\ell}
\R_{\vec\ell_1,\vec\ell}(-u)+
\R_{{\vec\ell}_2,{\vec\ell}_1}(u)\R^{\prime}_{{\vec\ell}_2,\vec\ell}(0)
\R_{\vec\ell_1,\vec\ell}(-u)=
$$
$$
=\R_{\vec\ell_1,\vec\ell}(-u)\R^{\prime}_{\vec\ell_2,\vec\ell}(0)
\R_{\vec\ell_2,\vec\ell_1}(u)+
\R_{\vec\ell_1,\vec\ell}(-u)\P_{\vec\ell_2,\vec\ell}
\R^{\prime}_{\vec\ell_2,\vec\ell_1}(u),
$$
then multiply both sides of the obtained equation by 
the permutation $\P_{\vec\ell_2,\vec\ell}$ from 
the left~($\vec\ell_2=~\vec\ell$):
$$
\R^{\prime}_{{\vec\ell},{\vec\ell}_1}(u)\R_{\vec\ell_1,\vec\ell}(-u)+
\R_{{\vec\ell},{\vec\ell}_1}(u)
\R^{\prime}_{{\vec\ell},{\vec\ell}_2}(0)\P_{\vec\ell_2,\vec\ell}
\R_{\vec\ell_1,\vec\ell}(-u)=
$$
$$
=\R_{\vec\ell_1,\vec\ell_2}(-u)
\P_{\vec\ell_2,\vec\ell}\R^{\prime}_{\vec\ell_2,\vec\ell}(0)
\R_{\vec\ell_2,\vec\ell_1}(u)+
\R_{\vec\ell_1,\vec\ell_2}(-u)\R^{\prime}_{\vec\ell_2,\vec\ell_1}(u),
$$
and calculate $\str_{V_{\vec\ell}}$ using the equalities:
$$
\str_{V_{\vec\ell}}
\R^{\prime}_{{\vec\ell},{\vec\ell}_1}(u)
\R_{\vec\ell_1,\vec\ell}(-u) = const
\ ;\ \str_{V_{\vec\ell}}\P_{\vec\ell_2,\vec\ell}
\R^{\prime}_{\vec\ell_2,\vec\ell}(0)=const.
$$
After all we obtain
$$
\str_{V_{\vec\ell}}\R_{{\vec\ell},{\vec\ell}_1}(u)
\R^{\prime}_{{\vec\ell},{\vec\ell}_2}(0)\P_{\vec\ell_2,\vec\ell}
\R_{\vec\ell_1,\vec\ell}(-u)=
\R_{\vec\ell_1,\vec\ell_2}(-u)
\R^{\prime}_{\vec\ell_2,\vec\ell_1}(u)
\cdot\str_{V_{\vec\ell}} 1 + const
$$
and using the evident identity~(consequence of eq.~(\ref{unitar})):
$$
\R_{\vec\ell_1,\vec\ell_2}(-u)
\R^{\prime}_{\vec\ell_2,\vec\ell_1}(u) = 
\R^{\prime}_{\vec\ell_1,\vec\ell_2}(-u)\R_{\vec\ell_2,\vec\ell_1}(u) 
+ const
$$
we arrive at the general formula: 
$$
\str_{V_{\vec\ell}}\R_{\vec\ell,\vec\ell_1}(u)
\H_{\vec\ell,\vec\ell_2}(0)\R_{\vec\ell_1,\vec\ell}(-u) =
\H_{\vec\ell_1,\vec\ell_2}(-u)\cdot \str_{V_{\vec\ell}} 1 + const
$$
Finally we obtain the following representation 
for $\TT^{\prime}_{\vec\ell}(0)$:
$$
\TT^{-1}_{\vec\ell}(0)
\TT^{\prime}_{\vec\ell}(0)= \H_{\vec\ell_1,\vec\ell_2}(-c_1)+
\sum_{i=3}^{N-1} \H_{\vec\ell_{i-1},\vec\ell_i}(0)+
\H_{\vec\ell_{N-1},\vec\ell_N}(-c_N)+ const
$$
This operator, commuting with all integrals of motions, is a sum of
two-particle operators and can be considered as the Hamiltonian.
The two-particle Hamiltonians entering the sum can 
be easily calculated from the universal R-matrix by~(\ref{Hop}).
The expression for the $\H_{\vec\ell_{i-1},\vec\ell_i}(0)$ 
coincides with the two-particle Hamiltonian for 
the periodic chain~(\ref{Hper}) up to overall coefficient two but 
the expression for the $\H_{\vec\ell_1,\vec\ell_2}(c)$ is 
rather lengthy to be presented here.

\section{Conclusions}

In this paper we have obtained the general 
solution of the Yang-Baxter equation acting on the tensor product 
of arbitrary representations of the superalgebra~$s\ell(2|1)$.

We have represented the lowest weight module of~$s\ell(2|1)$ 
by polynomials in one even~($z$) and two 
odd~($\theta,\bar\theta$) variables.
Therefore the general R-matrices is an operator acting 
on two-point functions being polynomials 
in the two sets of variables~($z_1,\theta_1,\bar\theta_1$) 
and~($z_2,\theta_2,\bar\theta_2$). 
Instead of calculating this operator explicitly 
we have obtained its matrix elements on the 
space of lowest weights. 
The eigenfunctions and eigenvalues can be easily 
calculated from the obtained matrix.\\
From the general R-matrix for two isomorphic 
representations we have calculated the nearest 
neighbour interaction Hamiltonians for an homogeneous closed chain.
The result applies both for the finite and 
infinite-dimensional representations on the sites.

Using the general R-matrix an integrable open chain can be 
constructed with arbitrary isomorphic representations 
on the inner sites and other arbitrary representations 
at the end points.
The nearest neighbour interaction Hamiltonian has been calculated.

\section{Acknowledgments}

We thank L.N.Lipatov and A.Manashov for the stimulating 
discussion and critical remarks.
This work has been supported by 
Deutsche Forschungsgemeinschaft, grant No  KI 623/1-2
and by INTAS,grant No 96-524. 
One of us (D.K.) is grateful to 
Saxonian Ministry of Science and Arts for supporting 
his visit at Leipzig University.

\section{Appendix A}

In this Appendix we discuss briefly the derivation 
of the expression~(\ref{solABCD}) for the general R-matrix. 

In matrix form the defining relation for the R-matrix reads as follows:
\be
\R_{\vec\ell_1\vec\ell_2}(u) K_{AB} = 
\bar{K}_{AB} \R_{\vec\ell_1\vec\ell_2}(u)
\ ;\  A,B = 1,2,3.
\label{infom}
\ee
where:
$$
K = \left (\begin{array}{ccc}
K_{11} & K_{12} & K_{13} \\
K_{21} & K_{22} & K_{23} \\
K_{31} & K_{32} & K_{33}
\end{array} \right )\equiv \frac{u}{2\eta}(\R_2-\R_1)+\R_1\R_2
\ ;\  \bar{K}\equiv \frac{u}{2\eta}(\R_2-\R_1)+\R_2\R_1.
$$
The matrix element $\bar{K}_{AB}$ can be obtained from 
$K_{AB}$ by formal substitution $1 \leftrightarrow 2$ and $u \leftrightarrow -u$:
$$
\ell_1,b_1,Z_1 \leftrightarrow \ell_2,b_2,Z_2
\ ;\ u \leftrightarrow -u , 
$$
where $Z\equiv (z, \theta, \bar\theta)$.\\
The operators $K_{AB}$ and lowest weights transform as follows: 
$$ 
K_{AB}\leftrightarrow\bar{K}_{AB}
\ ;\  \Phi^{\pm}_{n} \leftrightarrow (-1)^n\Phi^{\mp}_{n}
\ ;\  \Psi^{\pm}_{n} \leftrightarrow (-1)^{n+1}\Psi^{\pm}_{n}.
$$ 
There are nine equations and we start from the simplest one.

\subsection{Equation $\R K_{13}=\bar{K}_{13}\R$}
\setcounter{equation}{0}

The operator $K_{13}$ commutes with the lowering 
generators $V^{-},W^{-},S^{-}$ and 
the covariant derivatives $D^{\pm}_1$.
Therefore operator $K_{13}$ maps lowest weight vectors 
with definite chirality to lowest weight vector 
with the same chirality and decrease its power by one:  
$$
K_{13}\Phi^{\pm}_{n}=\alpha^{\pm}_{n}(\u)\Phi^{\pm}_{n-1}
\ ;\ K_{13}\Psi^{\pm}_{n}=\beta^{\pm}_{n}(\u)\Psi^{\pm}_{n-1}
$$ 
The explicit calculation gives:
$$
\alpha^{+}_{n}(\u) = n(\u+\ell_n)
\ ;\ \alpha^{-}_{n}(\u) = n (\u+\ell_n-1)
\ ;\ \beta^{\pm}_{n}(\u) = n (\u+\ell_n),
$$
where:
$$
\ell_n \equiv n+\ell_1+\ell_2
\ ;\ \u\equiv \frac{u}{\eta}+ b_1-b_2.
$$
The action of operator $\bar{K}_{13}$ on lowest weights 
vectors can be obtained from formulae for $K_{13}$ 
by formal substitution $\ell_1,b_1,Z_1 \leftrightarrow \ell_2,b_2,Z_2$ and 
$u \leftrightarrow -u$:
$$
\bar{K}_{13}\Phi^{\pm}_{n}=-\alpha^{\mp}_{n}(-\u)\Phi^{\pm}_{n-1}
\ ;\ K_{13}\Psi^{\pm}_{n}=-\beta^{\pm}_{n}(-\u)\Psi^{\pm}_{n-1}.
$$
We project the operator equation 
$\R K_{13}=\bar{K}_{13}\R$ onto the lowest weight 
vectors $\Phi^{\pm}_{n}$:
$$
\R K_{13}\Phi^{+}_{n} = \bar{K}_{13}\R \Phi^{+}_{n}
\Longrightarrow
$$
$$
\alpha^{+}_{n}(\u)[A_{n-1}\Phi^{+}_{n-1}+B_{n-1}\Phi^{-}_{n-1}]=
-A_{n}\alpha^{-}_{n}(-\u)\Phi^{+}_{n-1}-
B_{n}\alpha^{+}_{n}(-\u)\Phi^{-}_{n-1}
$$
$$
\R K_{13}\Phi^{-}_{n} =
\bar{K}_{13}\R \Phi^{-}_{n}
\Longrightarrow
$$
$$
\alpha^{-}_{n}(\u)[C_{n-1}\Phi^{+}_{n-1}+D_{n-1}\Phi^{-}_{n-1}]=
-C_{n}\alpha^{-}_{n}(-\u)\Phi^{+}_{n-1}-
D_{n}\alpha^{+}_{n}(-\u)\Phi^{-}_{n-1}
$$
which results in the recurrent relations:
$$
\alpha^{+}_{n}(\u)A_{n-1}=
-\alpha^{-}_{n}(-\u)A_{n}
\ ;\ \alpha^{+}_{n}(\u) B_{n-1}= 
-\alpha^{+}_{n}(-\u)B_{n}
$$
$$
\alpha^{-}_{n}(\u)C_{n-1}=
-\alpha^{-}_{n}(-\u)C_{n}
\ ;\ \alpha^{-}_{n}(\u) D_{n-1}=
-\alpha^{+}_{n}(-\u)D_{n}
$$
with the following general solution:
$$
A_{n}(u)= A(-1)^{n+1}
\frac{\Gamma\left(\u+\ell_n+1\right)}
{\Gamma\left(-\u+\ell_n\right)}
\ ;\  B_{n}(u)=B (-1)^{n}
\frac{\Gamma\left(\u+\ell_n+1\right)}
{\Gamma\left(-\u+\ell_n+1\right)}
$$
$$
C_{n}(u)= C(-1)^{n}
\frac{\Gamma\left(\u+\ell_n\right)}
{\Gamma\left(-\u+\ell_n\right)}
\ ;\  D_{n}(u)= D (-1)^{n+1}
\frac{\Gamma\left(\u+\ell_n\right)}
{\Gamma\left(-\u+\ell_n+1\right)}
$$
The projection onto the odd lowest weight 
vectors $\Psi^{\pm}_{n}$ leads to analogous recurrent 
relations 
$$
\beta_{n}(\u)F_{n-1}=-\beta_{n}(-\u)F_{n}
\ ;\ \beta_{n}(\u)E_{n-1}=-\beta_{n}(-\u)E_{n}
$$
with general solution:
$$
E_{n}(u)= E(-1)^{n}
\frac{\Gamma\left(\u+\ell_n+1\right)}
{\Gamma\left(-\u+\ell_n+1\right)}
\ ;\  F_{n}(u)= F(-1)^{n}
\frac{\Gamma\left(\u+\ell_n+1\right)}
{\Gamma\left(-\u+\ell_n+1\right)}.
$$
We see that equation $\R K_{13}=\bar{K}_{13}\R$ 
fixes the n-dependence of the matrix elements of $\R$-matrix.
The remaining equations fix the coefficients 
$A,B,...$ up to overall normalization.

\subsection{Equation $\R K_{12} = \bar{K}_{12}\R $}
\setcounter{equation}{0}

Due to commutativity of the operator $K_{12}$ 
with lowering generators $W^{-},S^{-}$ 
we can write down the general formulae for the 
action of operator $K_{12}$ on even lowest weight vectors:  
$$
K_{12}\Phi^{\pm}_{n} = 
a^{\pm} W^{+}\Phi^{+}_{n-1}+ b^{\pm} W^{+}\Phi^{-}_{n-1} + 
c^{\pm}\Psi^{-}_{n-1}.
$$
The coefficients $a^{\pm}$ and $b^{\pm}$ can be 
calculated using the following commutation relation:  
$$
\{V^{-}, K_{12}\} = K_{13}.
$$
Remembering the known results for operator $K_{13}$
$$
\{V^{-}, K_{12}\} = K_{13} \Rightarrow 
V^{-}K_{12}\Phi^{\pm}_{n}= K_{13}\Phi^{\pm}_{n}=
\alpha^{\pm}_n(\u)\Phi^{\pm}_{n-1}
$$
and the simple formula~($b=b_1+b_2$):
$$
V^{-} W^{+}\Phi^{\pm}_{n}=-(\ell_n+b) \Phi^{\pm}_{n}\Rightarrow 
V^{-}K_{12}\Phi^{\pm}_{n}= 
-a^{\pm}(\ell_{n-1}+b)\Phi^{+}_{n-1} - 
b^{\pm}(\ell_{n-1}+b)\Phi^{-}_{n-1},
$$
we obtain:
$$ 
(\ell_{n-1}+b) a^{+}=-\alpha^{+}_n(\u) \ ,\  b^{+} = 0
\ ;\  a^{-}= 0
\ ,\  (\ell_{n-1}+b) b^{-} = -\alpha^{-}_n(\u).
$$
The coefficient $c^{+}$ can be 
calculated using the commutativity 
of $K_{12}$ and $D^{+}_1$:  
$$
\{D^{+}_1, K_{12}\} = 0 
\Rightarrow D^{+}_1 K_{12}\Phi^{+}_{n} = 
a^{+}D^{+}_1 W^{+}\Phi^{+}_{n-1}+c^{+}D^{+}_1\Psi^{-}_{n-1}= 0
$$
and the simple formulae:
$$
D^{+}_1 W^{+}\Phi^{+}_{n-1} = (\ell_1+b_1)\Phi^{+}_{n-1}
\ ;\ D^{+}_1 \Psi^{-}_{n-1} = - \Phi^{+}_{n-1}.
$$
We obtain
$$
 (\ell_1+b_1)a^{+} = c^{+}.
$$
The coefficient $c^{-}$ can be 
calculated using the commutation relation with 
the covariant derivative $D^{+}_2$:  
$$
\{D^{+}_2, K_{12}\} = D^{+}_2 W^{-}_1 + (\ell_2+b_2)S^{-}_1
\Rightarrow 
$$
$$
D^{+}_2 K_{12}\Phi^{-}_{n} = 
b^{-}D^{+}_2 W^{+}\Phi^{-}_{n-1}+c^{-}D^{+}_2\Psi^{-}_{n-1}=
(\ell_2+b_2)S^{-}_1\Phi^{-}_{n}
$$
and the formulae:
$$
D^{+}_2 W^{+}\Phi^{-}_{n-1} = (\ell_2+b_2)\Phi^{-}_{n-1}
\ ;\ D^{+}_2 \Psi^{-}_{n-1} = \Phi^{+}_{n-1}
\ ;\ S^{-}_1 \Phi^{-}_{n} = -n\Phi^{-}_{n-1}.
$$
We obtain
$$
(\ell_2+b_2)b^{-}+c^{-}=-n(\ell_2+b_2).
$$
Finally we have~($b=b_1+b_2$):
$$
K_{12}\Phi^{+}_{n} = 
-\frac{\alpha^{+}_n(\u)}{\ell_{n-1}+b} W^{+}\Phi^{+}_{n-1}- 
\frac{\alpha^{+}_n(\u)(\ell_1+b_1)}{\ell_{n-1}+b}\Psi^{-}_{n-1}
$$
$$
K_{12}\Phi^{-}_{n} = 
-\frac{\alpha^{-}_n(\u)}{\ell_{n-1}+b}W^{+}\Phi^{-}_{n-1}+
\frac{n(\u-b_1-b_2)(\ell_2+b_2)}{\ell_{n-1}+b}
\Psi^{-}_{n-1}.
$$
The analogous calculations for odd lowest weight vectors give:
$$
K_{12}\Psi^{+}_{n} = 
-\frac{\beta_n(\u)}{\ell_{n}+b} W^{+}\Psi^{+}_{n-1}+
\frac{(\u-b)(\ell_2+b_2)}{\ell_{n}+b}\Phi^{+}_{n} + 
\frac{\beta_n(\u)(\ell_1+b_1)}{n(\ell_{n}+b)}\Phi^{-}_{n} 
\ ;\  K_{12}\Psi^{-}_{n} = -
\frac{\beta_n(\u)}{\ell_{n-1}+b} W^{+}\Psi^{-}_{n-1}
$$
The expression for the action of the operator 
$\bar{K}_{12}$ is obtained by symmetry:
$$
\bar{K}_{12}\Phi^{-}_{n} = 
\frac{\alpha^{+}_n(-\u)}{\ell_{n-1}+b} W^{+}\Phi^{-}_{n-1}- 
\frac{\alpha^{+}_n(-\u)(\ell_2+b_2)}{\ell_{n-1}+b}\Psi^{-}_{n-1}
$$
$$
\bar{K}_{12}\Phi^{+}_{n} = 
\frac{\alpha^{-}_n(-\u)}{\ell_{n-1}+b}W^{+}\Phi^{+}_{n-1}-
\frac{n(\u+b)(\ell_1+b_1)}{\ell_{n-1}+b}
\Psi^{-}_{n-1}
$$
$$
\bar{K}_{12}\Psi^{+}_{n} = 
\frac{\beta_n(-\u)}{\ell_{n}+b} W^{+}\Psi^{+}_{n-1}+
\frac{(\u+b)(\ell_1+b_1)}{\ell_{n}+b}\Phi^{-}_{n}- 
\frac{\beta_n(-\u)(\ell_2+b_2)}{n(\ell_{n}+b)}\Phi^{-}_{n} 
\ ;\  \bar{K}_{12}\Psi^{-}_{n} =
\frac{\beta_n(-\u)}{\ell_{n-1}+b} W^{+}\Psi^{-}_{n-1}.
$$
The projection of the operator equation 
$\R K_{12}= \bar{K}_{12}\R$ onto lowest weight 
vectors leads to the following new recurrent relations~($b=b_1+b_2$):
\be
\R K_{12}\Phi^{+}_{n} = \bar{K}_{12}\R \Phi^{+}_{n}
\Longrightarrow
\label{k12}
\ee
$$
(\ell_1+b_1)\alpha^{+}_n(\u)E_{n-1} = 
n(\ell_1+b_1)(\u+b)A_n+(\ell_2+b_2)\alpha^{+}_n(-\u)B_{n}
$$
$$
\R K_{12}\Phi^{-}_{n} = \bar{K}_{12}\R \Phi^{-}_{n}
\Longrightarrow
$$
$$
-n(\ell_2+b_2)(\u-b)E_{n-1} = 
n(\ell_1+b_1)(\u+b_1+b_2)C_n+(\ell_2+b_2)\alpha^{+}_n(-\u)D_{n}
$$
$$
\R K_{13}\Phi^{+}_{n} = \bar{K}_{13}\R \Phi^{+}_{n}
\Longrightarrow
$$
$$
n(\ell_2+b_2)(\u-b)A_n+(\ell_1+b_1)\beta_n(\u)C_{n}
=-(\ell_2+b_2)\beta_n(-\u)F_{n}
$$
$$
n(\ell_2+b_2)(\u-b)B_n+(\ell_1+b_1)\beta_n(\u)D_{n}
=n(\ell_1+b_1)(\u+b_1+b_2)F_{n}
$$

\subsection{Equation $\R K_{23} = \bar{K}_{23}\R $}
\setcounter{equation}{0}

There exists an automorphism of the algebra $s\ell(2|1)$:
$$
W^{\pm} \leftrightarrow V^{\pm}
\ ;\ B \leftrightarrow -B
$$ 
which has the following form in our representation:
$$
\theta \leftrightarrow \bar\theta
\ ;\ b \leftrightarrow -b.
$$
Due to this automorphism the matrix $K_{AB}$ has definite 
symmetry properties with respect to the transformation: 
$$
\ell_1,z_1 \leftrightarrow \ell_2,z_2
\ ;\ \theta_1,b_1 \leftrightarrow -\bar\theta_2,-b_2 
$$
The operators $K_{12}$,$K_{23}$ and lowest weights transform as follows: 
$$ 
K_{12}\leftrightarrow K_{23}
\ ;\ \Phi^{\pm}_{n} \leftrightarrow (-1)^n\Phi^{\pm}_{n}
\ ;\  \Psi^{\pm}_{n} \leftrightarrow (-1)^{n+1}\Psi^{\mp}_{n}.
$$
This symmetry allows to use the results of 
previous section and write down the 
formulae for the action of the operator $K_{23}$ on 
lowest weight vectors~($b=b_1+b_2$):
$$
K_{23}\Phi^{+}_{n} = 
\frac{\alpha^{+}_n(\u)}{\ell_{n-1}-b} V^{+}\Phi^{+}_{n-1}- 
\frac{\alpha^{+}_n(\u)(\ell_2-b_2)}{\ell_{n-1}-b}\Psi^{+}_{n-1},
$$
$$
K_{23}\Phi^{-}_{n} = 
\frac{\alpha^{-}_n(\u)}{\ell_{n-1}-b}V^{+}\Phi^{-}_{n-1}+
\frac{n(\u+b)(\ell_1-b_1)}{\ell_{n-1}-b}
\Psi^{+}_{n-1},
$$
$$
K_{23}\Psi^{-}_{n} = 
\frac{\beta_n(\u)}{\ell_{n}-b} V^{+}\Psi^{-}_{n-1}-
\frac{(\u+b)(\ell_1-b_1)}{\ell_{n}-b}\Phi^{+}_{n}- 
\frac{\beta_n(\u)(\ell_2-b_2)}{n(\ell_{n}-b)}\Phi^{-}_{n} 
\ ;\  K_{23}\Psi^{+}_{n} = 
\frac{\beta_n(\u)}{\ell_{n-1}-b} V^{+}\Psi^{+}_{n-1},
$$
$$
\bar{K}_{23}\Phi^{-}_{n} = 
-\frac{\alpha^{+}_n(-\u)}{\ell_{n-1}-b} V^{+}\Phi^{-}_{n-1}- 
\frac{\alpha^{+}_n(-\u)(\ell_1-b_1)}{\ell_{n-1}-b}\Psi^{+}_{n-1},
$$
$$
\bar{K}_{23}\Phi^{+}_{n} = 
-\frac{\alpha^{-}_n(-\u)}{\ell_{n-1}-b}V^{+}\Phi^{+}_{n-1}-
\frac{n(\u-b)(\ell_2-b_2)}{\ell_{n-1}-b}
\Psi^{+}_{n-1},
$$
$$
\bar{K}_{23}\Psi^{-}_{n} = 
-\frac{\beta_n(-\u)}{\ell_{n}-b} V^{+}\Psi^{-}_{n-1}-
\frac{(\u-b)(\ell_2-b_2)}{\ell_{n}-b}\Phi^{-}_{n}+ 
\frac{\beta_n(-\u)(\ell_1-b_1)}{n(\ell_{n}-b)}\Phi^{+}_{n} 
\ ;\  \bar{K}_{23}\Psi^{+}_{n} =
-\frac{\beta_n(-\u)}{\ell_{n-1}-b} V^{+}\Psi^{+}_{n-1}.
$$
The projection of the operator equation 
$\R K_{23}= \bar{K}_{23}\R$ onto lowest weight 
vectors leads to the following new recurrent relations~($b=b_1+b_2$):
\be
\R K_{23}\Phi^{+}_{n} = \bar{K}_{23}\R \Phi^{+}_{n}
\Longrightarrow
\label{k23}
\ee
$$
(\ell_2-b_2)\alpha^{+}_n(\u)F_{n-1} = 
n(\ell_2-b_2)(\u-b)A_n+(\ell_1-b_1)\alpha^{+}_n(-\u)B_{n}
$$
$$
\R K_{23}\Phi^{-}_{n} = \bar{K}_{23}\R \Phi^{-}_{n}
\Longrightarrow
$$
$$
-n(\ell_1-b_1)(\u+b)F_{n-1} = 
n(\ell_2-b_2)(\u-b)C_n+(\ell_1-b_1)\alpha^{+}_n(-\u)D_{n}
$$
$$
\R K_{23}\Phi^{+}_{n} = \bar{K}_{23}\R \Phi^{+}_{n}
\Longrightarrow
$$
$$
n(\ell_1-b_1)(\u+b)A_n+(\ell_2-b_2)\beta_n(\u)C_{n}
=-(\ell_1-b_1)\beta_n(-\u)E_{n}
$$
$$
n(\ell_1-b_1)(\u+b)B_n+(\ell_2-b_2)\beta_n(\u)D_{n}
=n(\ell_2-b_2)(\u-b)E_{n}
$$
The systems of equations~(\ref{k12}),~(\ref{k12}) 
fix the coefficients $A,B,C...$ up two arbitrary constants.
The next operator equation $\R K_{11} = \bar{K}_{11}\R$ 
fixes the remaining ambiguity but we avoid presenting 
the rather lengthy formulae here.   
Alternatively the missing equation can be obtained as follows.
The even lowest weight vectors $\Phi^{\pm}_{n}$ coincide for $n=0$.
Therefore one obtains: 
$$
\R \Phi^{+}_{0}=\R \Phi^{-}_{0} \Rightarrow
-A(\u+\ell_0)+B = \frac{C(-\u+\ell_0)}{\u+\ell_0}- 
\frac{D}{\u+\ell_0}
$$
Finally the systems~(\ref{k12}),~(\ref{k23}) 
and this last equation fix the solution completely:
$$
A=\u+b_{1}-b_{2}
\ ;\  B = (\ell_1+b_1)(\ell_2-b_2)
\ ;\  C = (\ell_1-b_1)(\ell_2+b_2)
$$
$$
D = (\ell_2-b_2)(\ell_2+b_2)\left(\u-b_{1}-b_{2}\right)-
\left(\u+b_{1}+b_{2}\right)
\left(\u-b_2-\ell_1\right)\left(\u-b_2+\ell_1\right)
$$
$$
E = (\u+b_1-\ell_2)(\u+b_1+\ell_2)
\ ;\  F = (\u-b_2-\ell_1)(\u-b_2+\ell_1).
$$
We have checked that obtained $\R$-matrix really is 
the solution also of the remaining seven equations but 
the involved formulae become rather lengthy starting 
from the equation $\R K_{11} = \bar{K}_{11}\R$.

\section{Appendix B}

The R-matrix acting in the tensor product of chiral 
modules can be obtained by simple reduction from  
the general R-matrix~(\ref{ABCD}).
First of all we have to change the overall 
normalization multiplying all matrix elements 
by the factor $(\ell_1-b_1)(\ell_2+b_2)$.
Let us consider all possible special cases:\\
{\bf 1.} chiral at 1, generic at 2 
$$
V_{\ell_1,-\ell_1}\otimes V_{\ell_2,b_2} = 
\sum_{n=0}^{\infty} V_{\ell+n,b}+
\sum_{n=0}^{\infty} V_{\ell+n+\half,b-\half}  
\ ;\ \ell_2\ne\pm b_2
\ ;\ \Phi^{+}_{n}\to V_{\ell+n,b}
\ ;\ \Psi^{+}_{n}\to V_{\ell+n+\half,b-\half}
$$
$$
\R_{\vec\ell_1\vec\ell_2}(u)
\left (\begin{array}{c}
\Phi^{+}_{n} \\ \Psi^{+}_{n} 
\end{array} \right ) = R\cdot
\left (\begin{array}{cc}
A_{n}(u) & 0 \\
0 & F_{n}(u) 
\end{array} \right )
\left (\begin{array}{c}
\Phi^{+}_{n} \\ \Psi^{+}_{n} 
\end{array} \right )
$$
$$
A_{n}(u)=(-1)^{n+1}
\frac{\Gamma\left(\u+\ell_n+1\right)}
{\Gamma\left(-\u+\ell_n\right)}
\ ;\  F_{n}(u)=(-1)^{n}
\frac{\Gamma\left(\u+\ell_n+1\right)}
{\Gamma\left(-\u+\ell_n+1\right)}
\cdot(\u+\ell_1-b_2)
$$
{\bf 2.} antichiral at 1, generic at 2 
$$
V_{\ell_1,\ell_1}\otimes V_{\ell_2,b_2} = 
\sum_{n=0}^{\infty} V_{\ell+n,b}+
\sum_{n=0}^{\infty} V_{\ell+n+\half,b+\half}  
\ ;\ \ell_2\ne\pm b_2  
\ ;\ \Phi^{-}_{n}\to V_{\ell+n,b}
\ ;\ \Psi^{-}_{n}\to V_{\ell+n+\half,b+\half}
$$
$$
\R_{\vec\ell_1\vec\ell_2}(u)
\left (\begin{array}{c}
\Phi^{-}_{n} \\ \Psi^{-}_{n} 
\end{array} \right ) = R\cdot
\left (\begin{array}{cc}
D_{n}(u) & 0 \\
0 & E_{n}(u) 
\end{array} \right )
\left (\begin{array}{c}
\Phi^{-}_{n} \\ \Psi^{-}_{n} 
\end{array} \right )
$$
$$
D_{n}(u)=(-1)^{n+1}
\frac{\Gamma\left(\u+\ell_n\right)}
{\Gamma\left(-\u+\ell_n+1\right)}
\cdot(\u-\ell_1-b_2)
\ ;\ E_{n}(u)=(-1)^{n}
\frac{\Gamma\left(\u+\ell_n+1\right)}
{\Gamma\left(-\u+\ell_n+1\right)}
$$
{\bf 3.} chiral at 1, antichiral at 2 
$$
V_{\ell_1,-\ell_1}\otimes V_{\ell_2,\ell_2} = 
\sum_{n=0}^{\infty} V_{\ell+n,b}
\ ;\ \Phi^{+}_{n}\to V_{\ell+n,b}
$$
$$
\R_{\vec\ell_1\vec\ell_2}(u)\Phi^{+}_{n}= R\cdot
A_{n}(u)\Phi^{+}_{n}
\ ;\ A_{n}(u)=(-1)^{n}
\frac{\Gamma\left(\u+\ell_n+1\right)}
{\Gamma\left(-\u+\ell_n\right)}
$$
{\bf 4.} antichiral at 1, chiral at 2
$$
V_{\ell_1,\ell_1}\otimes V_{\ell_2,-\ell_2} = 
\sum_{n=0}^{\infty} V_{\ell+n,b}
\ ;\ \Phi^{-}_{n}\to V_{\ell+n,b}
$$
$$
\R_{\vec\ell_1\vec\ell_2}(u)\Phi^{-}_{n}= R\cdot
D_{n}(u)\Phi^{-}_{n}
\ ;\ D_{n}(u)=(-1)^{n}
\frac{\Gamma\left(\u+\ell_n\right)}
{\Gamma\left(-\u+\ell_n+1\right)}
$$
{\bf 5.} antichiral at 1 and 2
$$
V_{\ell_1,\ell_1}\otimes V_{\ell_2,\ell_2} = 
\sum_{n=0}^{\infty} V_{\ell+n+\half,b+\half}  
\ ;\ \Psi^{-}_{n}\to V_{\ell+n+\half,b+\half}
$$
$$
\R_{\vec\ell_1\vec\ell_2}(u)\Psi^{-}_{n}= R\cdot
E_{n}(u)\Psi^{-}_{n}
\ ;\ E_{n}(u)=(-1)^{n}
\frac{\Gamma\left(\u+\ell_n+1\right)}
{\Gamma\left(-\u+\ell_n+1\right)}
$$
{\bf 6.} chiral at 1 and 2
$$
V_{\ell_1,-\ell_1}\otimes V_{\ell_2,-\ell_2} = 
\sum_{n=0}^{\infty} V_{\ell+n+\half,b-\half}
\ ;\ \Psi^{+}_{n}\to V_{\ell+n+\half,b-\half}  
$$
$$
\R_{\vec\ell_1\vec\ell_2}(u)\Psi^{+}_{n}= R\cdot
F_{n}(u)\Psi^{+}_{n}
\ ;\ F_{n}(u)=(-1)^{n}
\frac{\Gamma\left(\u+\ell_n+1\right)}
{\Gamma\left(-\u+\ell_n+1\right)}
$$
Note that we do not fix the overall normalization 
of the obtained R-matrix.

\section{Appendix C}
    
In this Appendix we discuss shortly the 
case of finite-dimensional representations and show that 
the obtained general R-matrix reduces 
to the known ones~\cite{PF} for 
the tensor product of modules with minimal dimensions.  

The tensor product of two finite-dimensional 
$s\ell(2|1)$-modules has the following 
direct sum decomposition~\cite{Mar}:
\be
V_{\ell_1,b_1}\otimes V_{\ell_2,b_2} = 
V_{\ell,b}+ V_{\ell+N,b}+
2\sum_{n=1}^{N-1} V_{\ell+n,b}+
\sum_{n=0}^{N-1} V_{\ell+n+\half,b-\half}+
\sum_{n=0}^{N-1} V_{\ell+n+\half,b+\half}  
\ ;\ \ell_i\ne\pm b_i 
\label{sumfd}
\ee
$$
\ell_1 =-\frac{n_1}{2}\ ,\  n_1\geq 2
\ ;\ \ell_2 =-\frac{n_2}{2}\ ,\  n_2\geq 2
\ ;\  N\equiv min(n_1,n_2)
\ ;\  \ell = -\frac{n_1+n_2}{2}\ ,\ b = b_1+b_2
$$
Note that this formula is applicable in the 
generic situation 
$\ell_i\ne\pm b_i\ ,\ \ell_i\ne -1/2$.
The direct sum decomposition 
reduces for the module $V_{-\half,b}$:
$$
V_{\ell_1,b_1}\otimes V_{-\half,b_2} = 
V_{\ell,b}+V_{\ell+1,b}+
V_{\ell+\half,b-\half}+V_{\ell+\half,b+\half}  
\ ;\ \ell =-\frac{n_1+1}{2} \ ,\ b = b_1+b_2
\ ;\  n_1\geq 2\ ,\ \ell_2 = -\half
$$
$$
V_{-\frac{1}{2},b_1}\otimes V_{-\frac{1}{2},b_2} = 
V_{-1,b}+V_{-\half,b-\half}+V_{-\half,b+\half}
\ ;\  b = b_1+b_2
\ ,\ \ell_1 =\ell_2= -\half.
$$
The origin of modifications for the tensor product involving the 
$V_{-\frac{1}{2},b}$-module is very simple.
The module $V_{-\frac{1}{2},b}$ is the four-dimensional 
vector space with the following basis:
$$
\Phi_0=1\ ;\ \Phi_1 = z+b\theta\bar\theta
\ ;\ \Psi^{+}_0 = \bar\theta\ ,\ \Psi^{-}_0 = \theta.
$$
It is evident that we are able to construct 
the following two-point lowest weight vectors only:
$$
V_{\ell_1,b_1}\otimes V_{-\half,b_2}\leftrightarrow
\Phi^{+}_{0}=\Phi^{-}_{0}= 1
\ ;\ (b_2-\half)\Phi^{+}_{0}-(b_2+\half)\Phi^{-}_{0}
\ ;\ \Psi^{+}_{0}=\bar\theta_{12}\ ;\ \Psi^{-}_{0}=\theta_{12}.
$$
Note that 
$$
(b_2-\half)\Phi^{+}_{0}-(b_2+\half)\Phi^{-}_{0}=
-z_1+b_2\theta_1\bar\theta_1 +z_2+b_2\theta_2\bar\theta_2+
(b_2-\half)\bar\theta_1\theta_2-(b_2+\half)\theta_1\bar\theta_2,
$$
$$
V_{-\frac{1}{2},b_1}\otimes V_{-\frac{1}{2},b_2}\leftrightarrow
\Phi^{+}_{0}=\Phi^{-}_{0}= 1
\ ;\ \Psi^{+}_{0}=\bar\theta_{12}
\ ;\ \Psi^{-}_{0}=\theta_{12}.
$$
Let us consider the general R-matrix~(\ref{ABCD}) 
acting in the tensor product 
$V_{-\frac{1}{2},b_1}\otimes V_{-\frac{1}{2},b_2}$.
In this case we have:
$$
\R_{\vec\ell_1\vec\ell_2}(u)= 
(A_0(u)+B_0(u))\cdot \P_1 +
E_0(u)\cdot \P_2 + F_0(u)\cdot \P_3
$$
where $\P_i$ are projectors on the 
modules in the direct sum decomposition: 
$$
V_{-\frac{1}{2},b_1}\otimes V_{-\frac{1}{2},b_2} = 
V_{-1,b}+V_{-\half,b-\half}+V_{-\half,b+\half}
\ ;\ b = b_1+b_2
\ ,\ \ell_1 =\ell_2= -\half
$$
$$
\P_1\rightarrow V_{-1,b}
\ ;\ \P_2\rightarrow V_{-\half,b-\half}
\ ;\ \P_3\rightarrow V_{-\half,b+\half}.
$$
After a simple calculation one obtains:
$$
\R_{\vec\ell_1\vec\ell_2}(u)\sim
\P_1 +
\frac{2\u-1+2b_1}{2\u+1-2b_2}\cdot \P_2 + 
\frac{2\u-1-2b_2}{2\u+1+2b_1}\cdot \P_3.
$$
This result coincides with the expression 
for the R-matrix given in~\cite{PF}:
$$
\R(\mu)= \P_1 +
\frac{4\mu-1+b_1+b_2}{4\mu+1-b_1-b_2}\cdot \P_2 +
\frac{4\mu-1-b_1-b_2}{4\mu+1+b_1+b_2}\cdot \P_3
$$
up to  the overall normalization and a redefinition 
of the spectral parameter:
$$
\u = 2\mu -\frac{b_1-b_2}{2}.
$$
The direct sum decomposition for the tensor product of 
the chiral modules (atypical representations) 
is well known~\cite{Mar}.
Now we need the simplest ones:
$$
V_{-\half,\half}\otimes V_{-\half,b_2} =  
V_{-1,b}+V_{-\half,b-\half}
\ ;\ b = b_1+b_2
\ ,\ \ell_1 =\ell_2= -\half\ ,\ b_1=\half
$$
Let us consider the R-matrix~(\ref{ABCD}) 
acting in the tensor product 
$V_{-\half,\half}\otimes V_{-\half,b_2}$.
In this case we have~(see Appendix B):
$$
\R_{\vec\ell_1\vec\ell_2}(u)= 
A_0(u)\cdot \P_1 + F_0(u)\cdot \P_2
$$
where $\P_i$ are projectors on the 
modules in the direct sum decomposition: 
$$
\P_1\rightarrow V_{-1,b}
\ ;\ \P_2\rightarrow V_{-\half,b-\half}.
$$
After a simple calculation one obtains:
$$
\R_{\vec\ell_1\vec\ell_2}(u)\sim 
\P_1 + \frac{2\u-1-2b_2}{2\u+2}\cdot \P_2.
$$
This result also coincides with the expression 
for such a $\R$-matrix given in~\cite{PF}.


\begin{thebibliography}{99}
\bibitem{SNR}
M.Scheunert, W.Nahm and V.Rittenberg ,
J.Math.Phys.{\bf 18} (1977) 155
\bibitem{JG}
P.D.Jarvis, H.S.Green ,
J.Math.Phys.{\bf 20} (1979) 2115
\bibitem{Mar}
M.Marcu,
J.Math.Phys.{\bf 21} (1980) 1277~,~
J.Math.Phys.{\bf 21} (1980) 1284
\bibitem{DIC}
L. Frappat, P. Sorba, A. Sciarrino,
{\it DICTIONARY ON LIE SUPERALGEBRAS,}\\
hep-th/9607161 
\bibitem{Maas}
Z.Maassarani,J.Phys.A: {\bf 28} (1995) 1305
\bibitem{RM}
P.B. Ramos, M.J. Martins, 
Nucl.Phys.{\bf B474}, (1996) 678 
\bibitem{EK}
F.E\ss ler, V.Korepin ,
Phys. Rev. {\bf B 46}, (1992) 9147
\bibitem{FK}
A.Foerster, M.Karowski ,
Nucl.Phys. {\bf B 396}, (1993) 611
\bibitem{PF}
M.P.Pfannm\"uller and H.Frahm ,
Nucl.Phys. {\bf B 479}, (1996) 575
\bibitem{S}
E.K.Sklyanin ,
J.Phys. {\bf A 21} (1988) 2375
\bibitem{AJ}
N.Andrei and H.Johannesson,
Phys. Lett.{\bf A 100} (1984) 108
\bibitem{FPT}
H.Frahm, M.P.Pfannm\"uller and A.M.Tsvelik ,
Phys. Rev. Lett. {\bf 81}, (1998) 2116
\bibitem{L} 
L.N. Lipatov,~{\it High-energy asymptotics of
multicolor QCD and exactly solvable lattice models,}
~Padova preprint DFPD-93-TH-70B,
~JETP.Lett. {\bf 59}(1994)596.
\bibitem{FK} 
L.D.Faddeev and G.P.Korchemsky, Phys.Lett.{\bf B342}(1995)311.
\bibitem{BDM}
V.Braun, S.Derkachov and A.Manashov,
Phys.Rev.Lett. {\bf 81}, (1998) 2020 
V.Braun, S.Derkachov, G.Korchemsky and A.Manashov,
Nucl.Phys. {\bf B 553}, (1999) 355 
\bibitem{DKM}
S.Derkachov, G.Korchemsky and A.Manashov,~{\it Evolution 
equations for quark-gluon 
distributions in multi-color QCD and open spin chains,}
hep-ph/9909539
\bibitem{B}
A.Belitsky, hep-ph/9907420,hep-ph/9903512,
Phys.Lett. {\bf B 453}, (1999) 59 
\bibitem{KK}
D.Karakhanian and R.Kirschner,~{\it Conserved 
currents of the three-reggeon interaction,}
hep-th/9902147;~{\it High-energy scattering 
in gauge theories 
and integrable spin chains,}~hep-th/9902031,
Fortschr. Phys.{\bf 48}, (2000) 139    
\bibitem{BFLK} 
A.P.Bukhvostov, G.V.Frolov,L.N.Lipatov and E.A.Kuraev,
Nucl.Phys. {\bf B 258}, (1985) 601 
\bibitem{BM} 
A.Belitsky, D.M\"uller, A.Sch\"afer,{\it Implications 
of $N=1$ supersymmetry for QCD conformal operators,}
~hep-ph/9811484,~Phys.Lett. {\bf B 450}, (1999) 126
\bibitem{F}
H.Frahm,~{\em Doped Heisenberg chains: 
spin S generalizations of the supersymmetric t-J model ,} 
~cond-mat/9904157,~Nucl.Phys.{\bf B 559}, (1999) 613 
\bibitem{KRS}
P.P. Kulish, N.Yu.Reshetikhin and E.K.Sklyanin,
Lett.Math.Phys. {\bf 5} (1981) 393-403 ,\\
E.K.Sklyanin,{\em "Quantum Inverse Scattering Method.Selected Topics"},
in "Quantum Group and Quantum Integrable Systems"
(Nankai Lectures in Mathematical Physics),
ed. Mo-Lin Ge,Singapore:World Scientific,1992,pp.63-97;
hep-th/9211111
\bibitem{KS}
P.P. Kulish and E.K.Sklyanin ,
Zap.Nauchn.Sem. LOMI {\bf 95} (1980) 129
\bibitem{K}
P.P. Kulish ,
Zap.Nauchn.Sem. LOMI {\bf 145} (1985) 140 ,
J.Soviet. Math. {\bf 35} (1986) 1111,\\
{\em Yang-Baxter equation and reflection 
equations in integrable models,} 
~hep-th/9507070
\bibitem{H}
V.O.Tarasov,L.A.Takhtajan and L.D.Faddeev
Theor.Math.Phys.57(1983) 163-181 ,
L.D. Faddeev, Les-Houches lectures 1995, hep-th/9605187

\end{thebibliography}
\end{document}